\documentclass[a4paper,11pt]{article}
\usepackage{jheppub} 
\usepackage{lineno}

\usepackage[english]{babel}

\usepackage{amssymb}

\usepackage{amsmath}
\usepackage{graphicx}

\DeclareMathOperator{\tr}{tr}

\DeclareMathOperator{\re}{Re}

\title{\boldmath Topological observables and $\theta$ dependence in high temperature QCD from lattice simulations}

\author[a]{A.Yu.~Kotov,}
\author[b]{M.P.~Lombardo}
\author[]{and A.~Trunin}

\affiliation[a]{Jülich Supercomputing Centre, Forschungszentrum Jülich, D-52428 Jülich, Germany}
\affiliation[b]{INFN, Sezione di Firenze, 50019 Sesto Fiorentino (FI), Italy}

\emailAdd{a.kotov@fz-juelich.de}
\emailAdd{lombardo@fi.infn.it}
\emailAdd{amtrnn@gmail.com}

\abstract{We study 
topology  in Quantum Chromodynamics at high temperatures by means of lattice calculations.
Simulations are performed
with $N_f=2+1+1$ Wilson twisted mass fermions at maximal twist with physical quark masses, and temperatures   \\$T~\gtrsim~180$~MeV. The results
obtained with three lattice spacings ranging between $0.057$ and $0.080$ fm are extrapolated to the continuum limit. We compare  the results for the topological susceptibility 
obtained with the field-theoretic definition with those
obtained from an observable constructed with the disconnected part of the chiral susceptibility, and
we confirm their agreement -- within the largish errors --  on our range of temperatures.  We also study the topological charge distribution,  the next order cumulant $b_2$ and,
for the first time, the Free Energy as a function of the $\theta$ angle.  We find a rapid crossover to the Dilute Instanton Gas behaviour above $ T \simeq 300$~MeV for all the observables we have considered.}

\begin{document}
\maketitle
\flushbottom

\section{Introduction}
\label{sec:intro}

Topological properties of QCD are very interesting for many different reasons. It has been understood long time ago that nontrivial topological fluctuations are important for the resolution of the $U_A(1)$ axial problem \cite{Witten:1979vv,Veneziano:1979ec}. Closely related to it is the conjecture of QCD axion~\cite{Peccei:1977hh,Peccei:1977ur,Weinberg:1977ma,Wilczek:1977pj}~--- an unobserved hypothetical particle, which was proposed in order to solve the mystery of zero $\theta$-term in QCD. It was also suggested some time ago that topological fluctuations in QCD can lead to the generation of local CP- and P-odd domains in the heavy ion collisions, which in turn result in the emergence of non-dissipative anomalous currents \cite{Fukushima:2008xe}. A more detailed discussion on different aspects of QCD topology and $\theta$ vacuum  can be found, e.g. in \cite{Lu:2020rhp, Vicari:2008jw,
Lombardo:2020bvn}. 

It all makes the study of topological properties of QCD a very important and interesting problem. One of the particular quantities of interest is the topological susceptibility~$\chi$ in QCD at high temperature. At large enough temperatures one could expect that its behaviour is described by the so-called dilute instanton gas approximation (DIGA) \cite{Gross:1980br}, while deeply in the confinement phase it should approach a non-zero value,
$\chi^{1/4} = 75.5(5)$  MeV \cite{GrillidiCortona:2015jxo}, thus solving the $U_A(1)$ axial problem. Instanton gas at high temperatures is expected to lead to nontrivial chiral effects, pronounced in the spectrum of the Dirac operator \cite{Kovacs:2023vzi}. The behaviour of~$\chi$ around the phase transition as well as the approach to the DIGA behaviour:
\begin{equation}
    \chi(T) \propto {T^{4-\beta_1}} \prod_{i=1}^{N_l}\frac{m_i}{T}\propto T^{-4b} 
    \label{eq:diga}
\end{equation}
with $\beta_1 = 11 N_c/3 - 2 N_l/3$, and $N_c, N_l$ number of colors and light flavors, quark masses $m_i$, 
are not yet fully understood. Note that within DIGA temperature dependence of the susceptibility $\chi$ is described by the exponent $b$, which is equal to $b=1.92$ for $N_l=2$ flavours and $b=2$ for $N_l=3$ light flavours. Due to the intrinsically non-perturbative nature of QCD, one of the most reliable approaches to this problem is lattice calculation. However, lattice studies of different groups within different approaches \cite{Chen:2022fid,Athenodorou:2022aay,Petreczky:2016vrs,Borsanyi:2016ksw} still do not quantitatively agree with each other, requiring further study of this problem. 

Beyond the topological susceptibility,  
the $\theta$-dependence in QCD 
is relevant both for theoretical and phenomenological
reasons~\cite{Vicari:2008jw}.  The $\theta$-term and the related chiral phase of the quark masses are 
collected in the effective vacuum angle $\theta= \theta + \text{Arg\, det}\,M$, where $M$ is the quark
mass matrix, which violates CP-symmetry.  Exceptions are not only $\theta=0$ (or if one of the quark masses would be zero), but also  $\theta = \pi$: in these cases the CP symmetry is exact. However, at $\theta=\pi$ there may be degenerate vacuum states with a possible spontaneous breaking of CP symmetry \cite{Witten:1979vv,Witten:1980sp,Smilga:1998dh,Vonk:2019kwv}.  In addition, the $\theta$ dependence may be a probe of non-perturbative dynamics at finite temperature, including the fate of the CP breaking transition and its possible endpoint.  However so far this aspect has been investigated mostly in model studies \cite{Gaiotto:2017yup,Bigazzi:2015bna,Parnachev:2008fy,Zhitnitsky:2008ha}. 

The traditional approach to the study of the vacuum structure
of QCD consists in the computation of the cumulants $c_{2n}(T)$ of the distribution of the topological charge at $\theta=0$.  Formally, the Free Energy density can then be reconstructed according to
\begin{equation}
    f(\theta,T)=\sum_{n}c_{2n}(T)\frac{\theta^{2n}}{(2n)!}.
\end{equation}
On the lattice, 
in full QCD only the first cumulant $\chi(T)\equiv c_2(T)$ and the second cumulant $b_2(T)\equiv\frac{c_4(T)}{12c_2(T)}$ have been computed \cite{Bonati:2015vqz,Burger:2018fvb}, and we will
 present below our results for physical pion mass and Wilson fermions.  These are important probes of the dynamics,
but their knowledge does not allow a full reconstruction of the Free Energy. 
Analytic continuation is an option (we remind the reader that due to the sign problem a direct calculation at non-zero $\theta$ is not possible), but
only a limited set of results is available, mostly in Yang-Mills theories~\cite{Bonanno:2024ggk,DElia:2013uaf,Panagopoulos:2011rb}.

In this paper we will explore   a different approach to the Free Energy. Consider the Grand Canonical Partition Function (GCPF) $Z$ and the Free Energy density $f$:
\begin{equation}
Z(\theta, T) = \sum Z_Q(T) \, e^{i Q \theta}  = e^{-V f(\theta, T)},
\label{eq:can}
\end{equation}
where $Z_Q(T)$ is the partition function in the fixed topological sector with topological charge $Q$.
Rather than computing the cumulants, we will simply use the topological charge distribution
to estimate the ratio $Z_Q/Z_0$ as a function of temperature. In principle, this allows the reconstruction of 
$Z(\theta, T)$, hence of the Free Energy, Eq. \ref{eq:can}.

In this manuscript we present the results of our study of topological properties of QCD at high temperatures $T>180$~MeV, obtained with Wilson twisted mass fermions at maximal twist --- a fermionic discretization with good control of chiral properties~\cite{Frezzotti:2003ni,Shindler:2007vp}. We discuss the distribution of the topological charge $Q$, its first two cumulants -- the topological susceptibility $\chi$ and kurtosis $b_2$ --  and the Free Energy of  QCD at nonzero~$\theta$. 

The remainder of this paper is organised as follows. In Sec.~\ref{sec:simdetails} we discuss details of lattice discretization used in this paper. We present the observables under study in Sec.~\ref{sec:observables}, and in Sec.~\ref{sec:autocorrelation} we discuss  aspects of the Monte Carlo evolution. In Sec.~\ref{sec:resultsfermionic} we show the results obtained from fermionic observables, and the analysis of gluonic observables is presented in Sec.~\ref{sec:resultsgluonic}. Finally, we summarize  our results and conclude in Sec.~\ref{sec:conclusions}.

Preliminary results of this study were presented in \cite{Kotov:2021ujj}.

\section{Simulation details}
\label{sec:simdetails}
\subsection{Details of lattice formulation}

We perform our study using $N_f=2+1+1$ Wilson-clover twisted mass fermions at the maximal twist. The simulations are carried out for the physical values of pion mass $m_{\pi}\approx 135$~MeV, though in the isospin symmetric limit $m_u=m_d$. Heavier strange and charm quarks are also tuned to reproduce physical mass values. The fermionic parts of the action, $S_{f}^l$ for the light doublet and $S_{f}^h$ in the heavy sector, read correspondingly:
\begin{equation}
\label{eq:fermaction}
S_{f}^l=\sum_{x,y}\bar{\chi}_l(x)\left(\delta_{x,y}-\kappa D^W_{x,y}+\frac{i}{2}\kappa c_{SW}\sigma^{\mu\nu}F^{\mu\nu}\delta_{x,y}  +2i\kappa a \mu_l\gamma_5\tau_3\delta_{x,y}\right)\chi_l(y),
\end{equation}
\begin{displaymath}
S_{f}^h=\sum_{x,y}\bar{\chi}_h(x)\left(\delta_{x,y}-\kappa D^W_{x,y}+\frac{i}{2}\kappa c_{SW}\sigma^{\mu\nu}F^{\mu\nu}\delta_{x,y}  +2\kappa a\mu_{\delta}\tau_1\delta_{x,y}+2i\kappa a \mu_{\sigma}\gamma_5\tau_3\delta_{x,y}\right)\chi_h(y),
\end{displaymath}
where $D_W$ is the standard Wilson operator, $c_{SW}\sigma^{\mu\nu}F^{\mu\nu}$ is the Sheikoleslami--Wohlert clover improvement term. The hopping parameter $\kappa$ is tuned to the critical value $\kappa_c$, called maximal twist, which automatically provides $O(a)$ improvement for observables. Parameter~$\mu_l$ describes the masses of the light quarks, while parameters $\mu_{\delta}$ and $\mu_{\sigma}$ are tuned to reproduce the ratio of strange-to-charm quark masses and the $D_s$ mass-to-decay constant ratio. For gluonic fields we employ Iwasaki gauge action:
\begin{equation}
    S_g[U]=\beta\left(c_0\sum_P \left[1-\frac13\re\tr U_P\right]+c_1\sum_R \left[1-\frac13\re\tr U_R\right]\right),
\end{equation}
where we take the sum over all plaquettes $P$ and $1\times 2$ rectangles $R$,
$c_0=3.648$ and $c_1=-0.331$.

This combination of gluonic and fermionic actions corresponds to the setup used by ETM Collaboration for zero temperature studies and thus allows us to use their results for parameter tuning and for measurements of such observables as lattice spacing, pion mass and renormalization factors~\cite{Alexandrou:2018egz,ExtendedTwistedMass:2021qui,Finkenrath:2022eon,ExtendedTwistedMass:2022jpw}.

\subsection{Ensembles at the physical point}

In this work we employ several ensembles generated with the physical pion mass $m_{\pi}\approx135$~MeV. In our simulations we utilize the parameter tuning of the ETM Collaboration at zero temperature~\cite{Alexandrou:2018egz,ExtendedTwistedMass:2021qui,Finkenrath:2022eon,ExtendedTwistedMass:2022jpw}. Simulations have been carried out in a fixed scale approach, in which we keep the bare lattice parameters fixed and change the temperature by varying the number of lattice spacings~$N_t$ in temporal direction. For technical reasons we generated only ensembles with even values of $N_t$. In order to take the continuum limit we have prepared configurations at three values of the lattice spacing $a\in[0.057,0.080]$ fm. Mainly, our configurations have large spatial volume $L_s\approx 5-5.5$ fm. To check finite volume effects we also generated one ensemble at the coarsest lattice spacing with smaller spatial volume $L_s\approx3.8$ fm. Our ensembles are summarized in Tab.~\ref{tab:physpionens}. Ensemble B64 was used in our previous work \cite{Kotov:2021rah}, and we increased statistics for several temperatures. The other three ensembles B48, C80 and D96 are newly generated.
Generation of configurations was performed with the help of standard Hybrid Monte Carlo algorithm. We saved each $4$th molecular dynamics trajectory and accumulated $O(500-1000)$ configurations per each ensemble and temperature value.

\begin{table}[]
    \centering
    \begin{tabular}{c|c|c|c|c|c}
    Ensemble & Lattice spacing [fm] & $N_s$ & $L_s$ [fm] & Values of $N_t$ &  T [MeV] \\
    \hline
    B48 & 0.07957(13) & 48 & 3.82 & 4,6,8,10,12,14 & 180-620 \\
    B64 & 0.07957(13) & 64 & 5.09 & 4,6,8,10,12,14,16,18,20 & 120-620  \\
    C80 & 0.06821(13) & 80 & 5.46 & 4,6,8,10,12,14,16 & 180-720 \\
    D96 & 0.05692(12) & 96 & 5.46 & 4,6,8,10,12,14,16,18,20 & 170-870
     \end{tabular}
    \caption{Summary of ensembles used in the current work. Lattice spacings are taken from \cite{ExtendedTwistedMass:2022jpw}, where they were determined from the pion decay constant $f_{\pi}$. For each set of parameters, given by ensemble and temperature, statistics is  $O(500-1000)$ configurations.}
    \label{tab:physpionens}
\end{table}

\section{Observables}
\label{sec:observables}
\subsection{Chiral condensate}

Light chiral condensate $\langle\bar{\psi}\psi\rangle_l$ is the standard order parameter of the chiral phase transition, which can be defined as the mass derivative of the partition function:

\begin{equation}
    \langle\bar{\psi}\psi\rangle_l=\frac{T}{V}\frac{\partial \log Z}{\partial m_l}=\frac{T}{V}\langle \tr M_l^{-1}\rangle,
    \label{eq:chcond}
\end{equation}
where $M_l$ is the Dirac operator corresponding to 2 light flavours, defined in Eq.~(\ref{eq:fermaction}).
\subsection{Chiral susceptiblity}
Chiral susceptibilility is defined as a mass derivative of the chiral condensate:

\begin{equation}
    \chi_\text{chiral}=\frac{V}{T}\frac{\partial\langle\bar{\psi}\psi\rangle_l}{\partial m_l}.
    \label{eq:chsus}
\end{equation}
It consists of connected and disconnected parts:

\begin{equation}
\begin{split}
    \chi_\text{chiral}=\chi_{\mathrm{conn}}+\chi_{\mathrm{disc}},\\
    \chi_{\mathrm{disc}}=\frac{V}{T}(\langle(\bar{\psi}\psi)^2\rangle_l-\langle\bar{\psi}\psi\rangle_l^2)=\frac{V}{T}\left(\langle(\tr M_l^{-1})^2\rangle-\langle\tr M_l^{-1}\rangle^2\right),\\
    \chi_{\mathrm{conn}}=-\frac{T}{V}\langle\tr M_l^{-2}\rangle.
\end{split}
\label{eq:chi_chiral_def}
\end{equation}
 We estimate chiral condensate and chiral susceptibility using $N_{\mathrm{stoch}}=24$ stochastic sources following our previous paper \cite{Kotov:2021rah}. Using already generated data we repeated the analysis with smaller number of stochastic sources $N_{\mathrm{stoch}}<24$ and checked that for $N_{\mathrm{stoch}}=24$ statistical errors are dominated by the gauge noise, making further increase of the $N_{\mathrm{stoch}}$ unnecessary.

\subsection{Topological charge and susceptibility}

Topology on the lattice can be measured by several methods. 

The gluonic definition is based on continuum definition for topological charge density in pure gluonic Yang-Mills theory~\cite{Luscher:2010iy}:
\begin{equation}
\label{eq:q-dens}
q(x)=\frac1{32\pi^2}\varepsilon_{\mu\nu\rho\sigma} \tr [F^{\mu\nu}(x) F^{\rho\sigma}(x)],
\end{equation}
where $F^{\mu\nu}(x)$ is continuous field strength tensor. Then, the topological charge is defined as
\begin{equation}
\label{eq:top-charge}
Q=\int\! q(x)\,d^4x.
\end{equation}
The definitions~\eqref{eq:q-dens}--\eqref{eq:top-charge} are straightforwardly implemented on the lattice:
\begin{equation}
\label{eq:Q-lattice}
Q=\frac{a^4}{32\pi^2}\varepsilon_{\mu\nu\rho\sigma}\sum_n \tr [F_{\mathrm{lat}}^{\mu\nu}(n) F_{\mathrm{lat}}^{\rho\sigma}(n)],
\end{equation}
where summation is over all sites of the lattice. This naive transcription suffers from large lattice artifacts,
which make $Q$ non-integer, and many lattice studies have focused on  ways to control the ultraviolet fluctuations \cite{Borsanyi:2016ksw,Bonati:2014tqa,Burger:2018fvb,Petreczky:2016vrs,Athenodorou:2022aay}. Once this is done, one may compute the topological susceptibility as 
\begin{equation}
\label{eq:chi-def}
\chi =\frac{\langle Q^2\rangle}{V}, \qquad V=a^4N_s^3N_t.
\end{equation}

This method gives us the possibility to determine the topological charge $Q$ on each configuration and consequently study  the full distribution of the topological charge over different sectors $Z_Q/Z_0$, the Free Energy density at nonzero $\theta$, as well as the next order cumulant of the topological charge distribution $b_2$.
As mentioned, the gluonic definition of the topological charge is known to suffer from large cutoff effects, see
in particular, e.g. \cite{Athenodorou:2022aay,Bonati:2015vqz}, and some smoothing  of the gauge field is needed. In our study we use the gradient flow~\cite{Luscher:2009eq}. It was shown~\cite{Bonati:2014tqa,Alexandrou:2015yba,Alexandrou:2017hqw}, that in leading order of perturbation theory cooling, smearing and gradient flow lead to the identical result for an updated gauge variable, allowing to derive a relation between flow time~$\tau$ and number of cooling/smearing steps, so we are confident that our
gluonic results are independent on the smoothing techique we have chosen. 

 Within this method we first produce the smooth gauge field configurations using the gradient flow \cite{Luscher:2009eq}. On the smoothed configurations we measure the topological charge $Q=\int d^4x\, q(x)$. In order to reduce lattice artifacts and to make the values of the topological charge close to integer, we employ the standard Symanzik tree-level improved discretization of the topological charge density (also called 3-loop improved definition) \cite{deForcrand:1997esx}:
\begin{equation}
\begin{split}
    q^{\mathrm{impr}}(x)&=c_0 q^{\mathrm{clov}}(x) + c_1 q^{\mathrm{rect}}(x),\\
    c_1&=-\frac{1}{12}, \quad c_0=1-8c_1
\label{eq:qgluonicdef}
\end{split}
\end{equation}
with 
\begin{equation}
\begin{split}
    q^{\mathrm{clov}}(x) &= \frac{1}{32\pi^2}\epsilon_{\mu\nu\rho\sigma}\tr(C_{\mu\nu}^{\mathrm{plaq}}C_{\rho\sigma}^{\mathrm{plaq}}),\\ 
    q^{\mathrm{rect}}(x) &= \frac{2}{32\pi^2}\epsilon_{\mu\nu\rho\sigma}\tr(C_{\mu\nu}^{\mathrm{rect}}C_{\rho\sigma}^{\mathrm{rect}}).\\
    \end{split}
    \end{equation}
Here $C_{\mu\nu}^{\mathrm{plaq}}$ and $C_{\mu\nu}^{\mathrm{rect}}$ are the standard clover average over plaquettes and rectangles field strength tensors.

\begin{figure}
    \centering
    \includegraphics[width=0.6\linewidth]{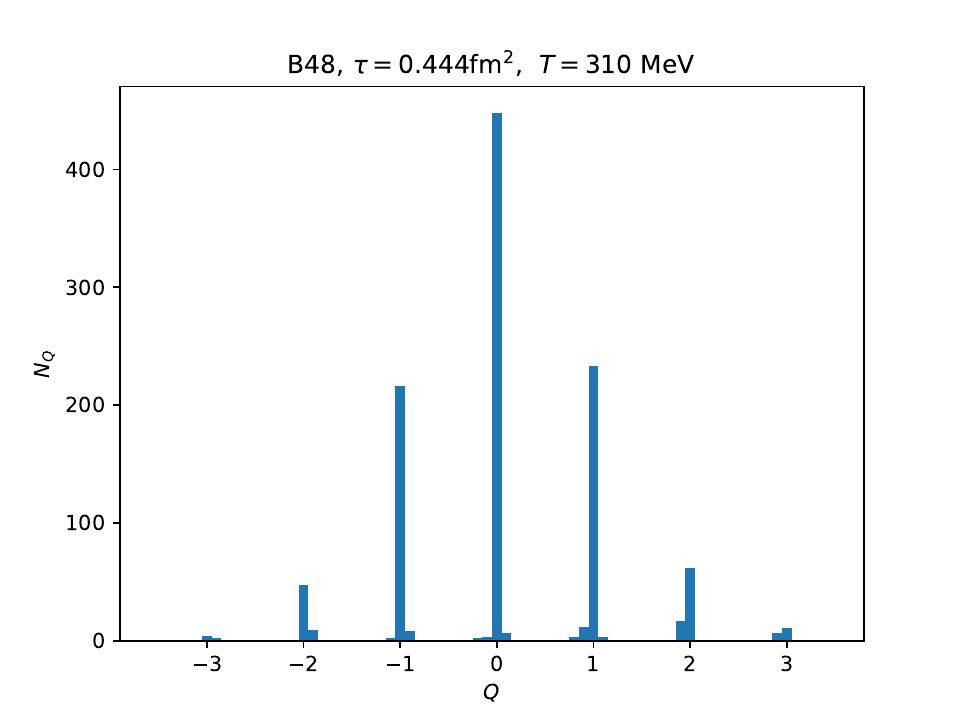}
    \caption{Histogram of the topological charge distribution for ensemble B48, $T\approx310$~MeV, gradient flow time $\tau=0.444$~fm$^2$.}
    \label{fig:histoundrounded}
\end{figure}

In Fig.~\ref{fig:histoundrounded} we present an example of histogram of the topological charge distribution for ensemble B64, $T\approx310$ MeV and large gradient flow time $\tau=0.444$~fm$^2$. One clearly sees peaks in the distribution, centered around near-integer values of $Q$. However, a close inspection of the histogram suggests that the exact position of the peaks is slightly shifted from integer numbers towards lower magnitude. Histograms for other ensembles, temperatures and flow times look similar and possess the same properties. In order to make values of the topological charge integer-valued, we employ the procedure proposed in \cite{DelDebbio:2002xa}. To this end we introduce a re-scaling factor $\alpha\sim1$ and define the rounded integer-values topological charge $Q_{\mathrm{round}}$ as

\begin{equation}
    Q_{\mathrm{round}}=[\alpha Q],
\end{equation}
where $[X]$ stands for the nearest integer to the real number $X$. The value of $\alpha$ is determined by minimizing the following functional~$F$:

\begin{equation}
    F=\sum_{\mathrm{conf}}(Q-[\alpha Q])^2,
\end{equation}
where the sum $\sum\limits_{\mathrm{conf}}$ is taken over the ensemble. In other words, we initially re-scale the topological charge by factor $\alpha$ with the purpose of bringing the peak positions closer to integer values and then round $Q$ to these integer values. The resulting re-scaled and rounded distribution of the one, shown in Fig.~\ref{fig:histoundrounded}, is presented in Fig.~\ref{fig:histo} (upper left plot).

\begin{figure}
    \centering
    \includegraphics[width=0.7\linewidth]{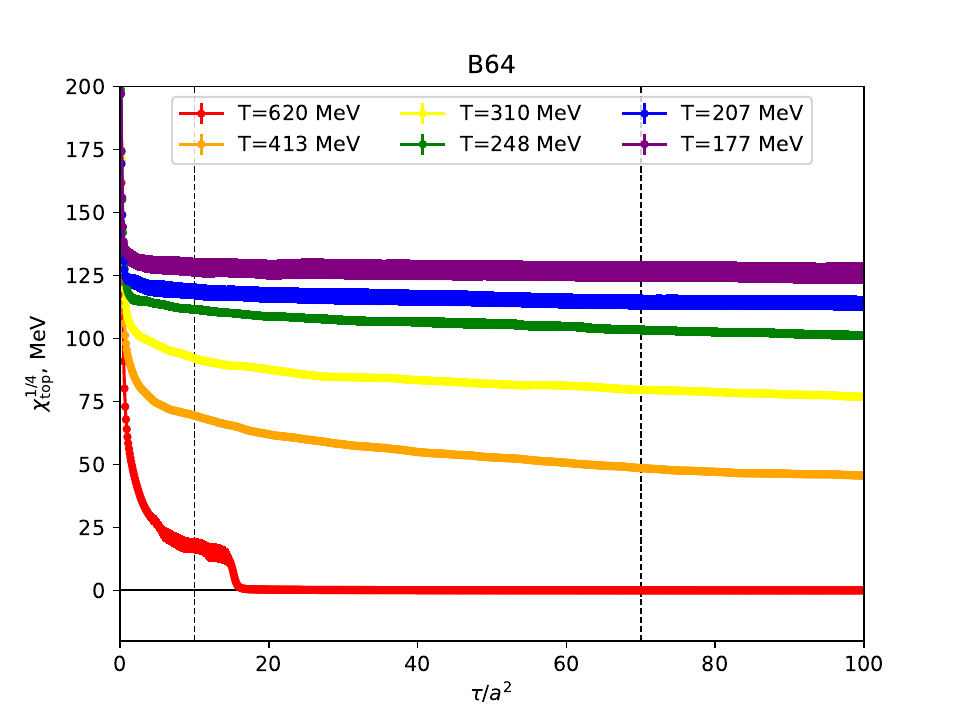}
    \caption{Topological susceptibility based on the gluonic definition as a function of gradient flow time $\tau$ for several temperatures. Data correspond to ensemble B64. For better readability we do not present data for low temperatures $T<177$ MeV, since they are very close to $T=177$ MeV data. Vertical dashed lines correspond to the GF times $\tau=0.063$ and $0.444$ fm$^2$ used in the analysis.}
    \label{fig:chivstaub}
\end{figure}

In this definition, the topological charge of each configuration $Q$ and consequently the topological susceptibility $\chi=\frac{\langle Q^2\rangle}{V}$ depend on the gradient flow time $\tau$. Typical dependence of the topological susceptibility on $\tau$ for several temperatures for ensemble B64 is presented in Fig.~\ref{fig:chivstaub}. At large gradient flow times $\tau\gtrsim a^2$ the topological charge $Q$ should become a constant function of $\tau$ \cite{Ce:2015qha}, and in agreement with this we observe  almost a horizontal plateau in the $\tau$-dependence of the topological susceptibility. In order to assess lattice cutoff effects, we study the continuum extrapolation of the topological susceptibility $\chi$ for two values of gradient flow time $\tau=0.063$ and $0.444$ fm$^2$. The former value $\tau=0.063$ fm$^2$ corresponds to the small flow time, for which we, however, already see clear integer-valued peaks in the histogram for the topological charge. The latter value $\tau=0.444$ fm$^2$ represents large flow times, where the $\tau$-dependency of the topological susceptibility has already reached a clear plateau. 

As an alternative way to determine the topological susceptibility, one may exploit the continuum relation between topological susceptibility and chiral density
$\bar\psi\gamma_5\psi$, as proposed in Ref.~\cite{Buchoff:2013nra}, and subsequently applied in Refs \cite{Petreczky:2016vrs,Kogut:1998rh,HotQCD:2012vvd}: 
\begin{equation}
Q=m_l\int d^4x\, \bar{\psi}(x)\gamma_5\psi(x),
\end{equation}
which is valid for any smooth gauge field configuration. As a consequence, the topological susceptibility $\chi=\frac{\langle Q^2\rangle}{V}$ is related to the disconnected $\gamma_5$ susceptibility: $\chi=m_l^2\chi_5^{\mathrm{disc}}$. 
In the chirally restored phase $\chi_5^{\mathrm{disc}}$  approximately equals the chiral disconnected susceptibility~\cite{HotQCD:2012vvd}. Since  
$\chi_5^{\mathrm{disc}}$ is a rather noisy quantity, one can take advantage of this relation to measure the
topological suceptibiliity from standard disconnected chiral susceptibility $\chi_{\mathrm{disc}}$~\eqref{eq:chi_chiral_def}, a much
easier quantity:
\begin{equation}
    \chi\approx m_l^2\chi_{\mathrm{disc}}.
\label{eq:topsusfermdef}
\end{equation}
In our calculation we take the light quark mass in the $\overline{MS}$ scheme at $\mu=2$~GeV \cite{FlavourLatticeAveragingGroupFLAG:2024oxs}:
\begin{equation}
m_l\left[~\overline{MS}(2~\mathrm{GeV})~\right]=3.427(51)~\mathrm{MeV}.
\label{eq:quarkmass}
\end{equation}

\subsection{Renormalization for Wilson twisted mass fermions}

The topological charge measured with gradient flow as described above does not require any additional renormalization \cite{Luscher:2010iy}. 

Fermionic method for the topological susceptibility~(\ref{eq:topsusfermdef}) uses the lattice results for the  disconnected chiral susceptibility. 
The chiral observables~\eqref{eq:chcond}--\eqref{eq:chsus} get additional renormalization.
The chiral condensate $\bar{\psi}\psi$ has both additive and multiplicative renormalization factors. Leading order additive divergence is proportional to $\mu/a^2$. In the disconnected chiral susceptibility additive divergences are subtracted out and thus are absent.   
Since combinations $m_l\bar{\psi}\psi$ and $m_l^2\chi$ (after subtracting the additive divergences) are RG-invariant \cite{Aoki:2006we}, the multiplicative renormalization factor for the chiral condensate is equal to the inverse mass renormalization factor $Z_m^{-1}$, and for the chiral susceptibility (both connected and disconnected) it equals $Z_m^{-2}$. $Z_m$ is related to the renormalization constant of the pseudoscalar density operator $Z_p$: $Z_m=Z_p^{-1}$~\cite{Frezzotti:2004wz}. The values of $Z_p$ for ensembles studied in the current work were calculated during zero temperature simulations by the ETM Collaboration in \cite{ExtendedTwistedMass:2022jpw}, and we present them in Tab.~\ref{tab:renormfactors}. In our analysis, in agreement with Eq.~(\ref{eq:quarkmass}), we use pseudoscalar renormalization factor $Z_p\left[~\overline{MS}(2~\mathrm{GeV})~\right]$ calculated in $\overline{MS}$-scheme at the scale $\mu=2$~GeV.

\begin{table}[ht]
\begin{center}
\begin{tabular}{|c|c|c|c|}
\hline
& $Z(B48/64)$, fm & $Z(C80)$, fm & $Z(D96)$, fm \\
\hline
$Z_p\left[~\overline{MS}(2~\mathrm{GeV})~\right]$ & 0.4788(54) & 0.4871(49) & 0.4894(44)  \\
\hline
\end{tabular}
\caption{Renormalization constants $Z_p$ for pseudoscalar bilinear in the $\overline{MS}$-scheme at $\mu=2$~GeV~\cite{ExtendedTwistedMass:2022jpw}.}
\label{tab:renormfactors}
\end{center}
\end{table} 

\section {Monte Carlo evolution}
\label{sec:autocorrelation}

\begin{figure}
    \centering
    \includegraphics[width=0.7\linewidth]{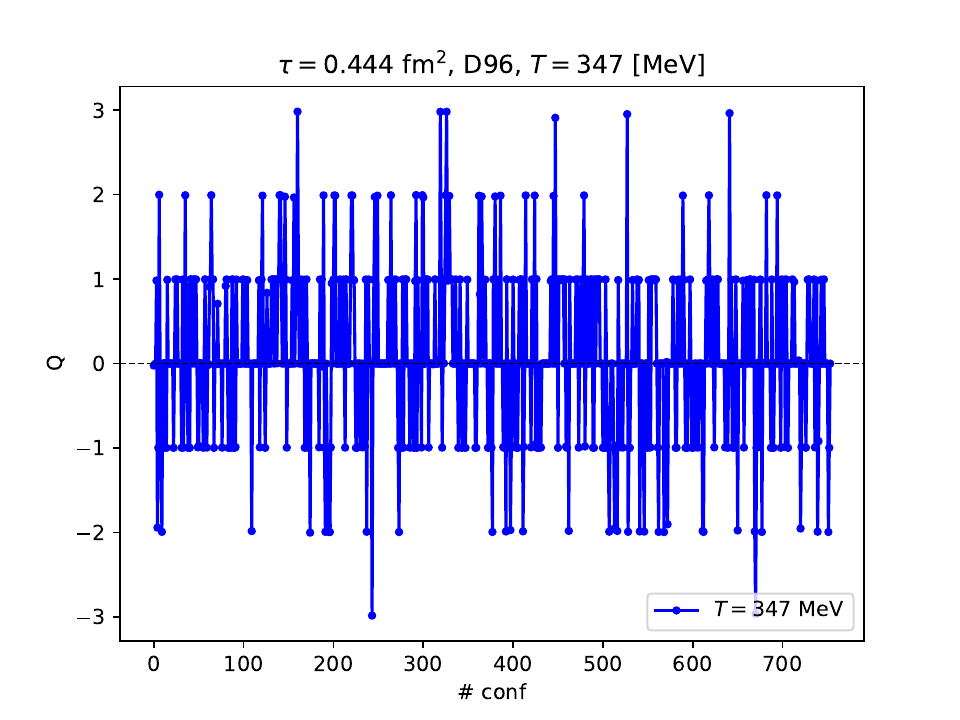}
    \caption{Monte Carlo time history of the gluonic definition of the topological charge $Q$ at GF time $\tau=0.444$ fm$^2$, ensemble D96, temporal lattice extent $N_t=10$ corresponds to~$T\approx347$~MeV.}
    \label{fig:q_d211_nt10}
\end{figure}

Markov chain Monte Carlo simulations are known to suffer from autocorrelation problems. One of the most problematic observables is the topological charge $Q$. It is an integer quantity, which separates the field space into different sectors, and tunnellings between different sectors are suppressed. The autocorrelation time of the topological charge rapidly grows with decreasing lattice spacing \cite{Schaefer:2010hu}, which potentially could lead to the system being stuck in one topological sector. We have checked that this is not the case in our simulations. As an example, in Fig.~\ref{fig:q_d211_nt10} we present Monte Carlo history of the gluonic definition of the topological charge on one of our ensembles with the finest latest spacing D96 and the temperature $T\approx 347$ MeV. It is determined using the gradient flow at the flow time $\tau=0.444$ fm$^2$. We clearly do not see the topological freezing and our simulations cover all relevant topological sectors. It is also worth noting that one clearly sees that topological charge takes only near-integer values for this ensemble. The plots for other lattice spacings and temperatures also do not show any topological freezing behaviour. We  measured the integrated autocorrelation time for different observables, and we have found that for all ensembles and all observables it is less than 10 configurations. In our analysis we split our data in 16 bins to take into account possible autocorrelations, the length of each bin is several times larger than autocorrelation length.

\section{Results on topological susceptibility from fermionic method}
\label{sec:resultsfermionic}

\begin{figure}
    \centering
    \includegraphics[width=0.9\linewidth]{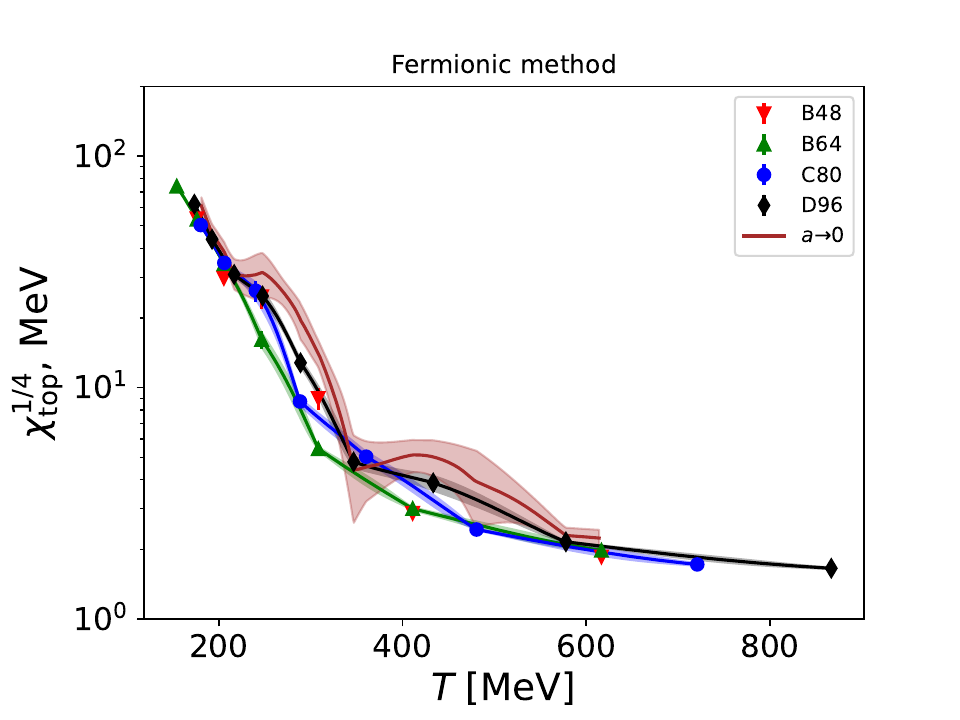}
    \caption{Temperature dependence of the topological susceptibility $\chi^{1/4}$ from fermionic method for all four ensembles under study and its continuum extrapolation.  Bands and lines represent the interpolation of our data described in the text.
    }
    \label{fig:chiferm}
\end{figure}

Fermionic method for measurements of the topological susceptibility is based on the Eq.~(\ref{eq:topsusfermdef}),
which relates the topological susceptibility in the chirally restored phase to the disconnected chiral susceptibility. 

In Fig.~\ref{fig:chiferm} we present the temperature dependence of the topological susceptibility for all four ensembles used in our simulations. Anticipating the discussion which we will make later, we also show the continuum extrapolated results. For the ensemble B we have collected results for two different spatial volumes. 
The finite size effects are below the statistical errors, with the exception of $T= 320$ MeV: it would be interesting to follow up on this and ascertain whether there are large size effects in this region, or if, instead, it is just a fluctuation. We will base our subsequent discussion on the results on the larger  lattices (B64).

\begin{figure}
\centering
\includegraphics[width=0.9\linewidth]
{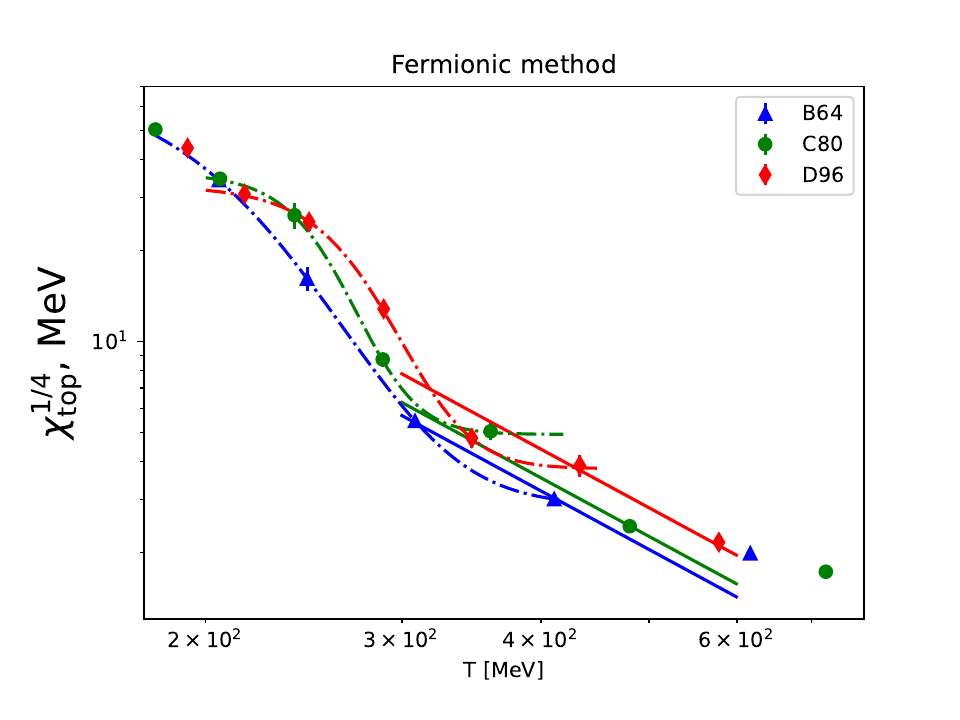}
\caption{The results on topological susceptibility on a double log scale.  We superimpose fits with the constrained exponent matching the DIGA value shown by solid lines. We also present fits with the logistic curve by dashed dotted lines in the crossover region around~$T\sim250$~MeV. 
}
\label{fig:comparediga}
\end{figure}

In Fig.~\ref{fig:comparediga} we plot the same results on a double log scale.
The plot shows  three different temperature regions: around the critical
temperatures, up to about 250 MeV; crossover region between 250 and 300
MeV; and a high temperature regime.  The straight lines shown in the plot 
are fits to the DIGA inspired form $\chi^{1/4}_{\mathrm{top}} \propto T^{-2}$, and at this stage are meant to guide the eye. Note that the $b = 2$ in Eq.~\eqref{eq:diga} corresponds to $N_l=3$, while for $N_l=2$, $b = 1.92$. The dashed dotted lines represent fits with the logistic curve in the crossover~region. 

\subsection{Continuum limit}

Since we perform the simulations in the fixed scale approach, i.e. only for fixed temperature values for each ensemble, we need to interpolate the data for intermediate temperature values in order to perform the continuum extrapolations at fixed temperature\footnote{As an alternative, one can try to fit data at a fixed lattice spacing with a physically motivated ansatz and perform the continuum extrapolation of the parameters of this ansatz. However, we have found, that it is a nontrivial task to find a suitable ansatz, which works in a wide range of temperatures and gives a reliable continuum extrapolation with small errors, except for DIGA regime $T\in [350:600]$~MeV, see Sec.~\ref{sec:diga}.}.  We used two different ways to interpolate: a simple linear interpolation between two consecutive points and a monotonic cubic interpolation, with the difference between two methods included in the systematic uncertainty. Results of the interpolation (with both systematic and statistical uncertainties) are presented by bands of the corresponding color in Fig.~\ref{fig:chiferm}. 

In Fig.~\ref{fig:chiferm} one can clearly see an interesting feature of the data: for low temperatures $T\lesssim 250$~MeV and for high temperatures $T\gtrsim 500$~MeV the results for all ensembles lie (almost) on one single curve. However, in the intermediate region $250\mathrm{~MeV}\lesssim T \lesssim 500\mathrm{~MeV}$ there are clear discrepancies among  different ensembles, what makes the continuum extrapolation in this region difficult.

\begin{figure}
    \centering
    \includegraphics[width=0.9\linewidth]{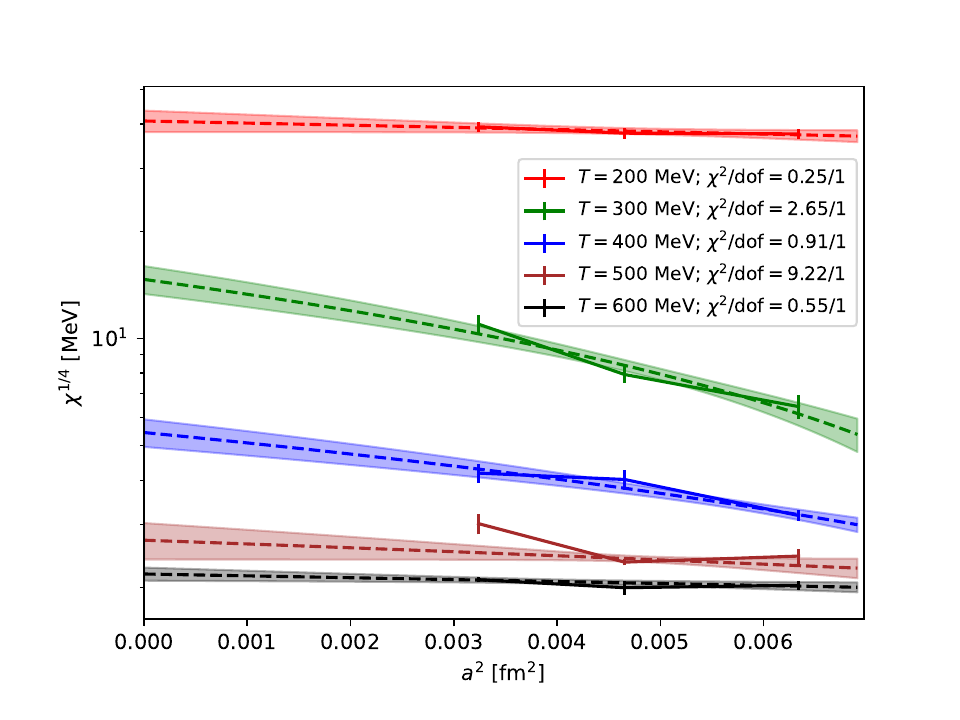}
    \caption{Topological susceptibility $\chi^{1/4}$ as a function of the squared lattice spacing $a^2$ for several temperatures. Bands represent a simple $a^2$ continuum limit: $\chi^{1/4}(a^2)=\chi^{1/4}(a=0)+Aa^2$.}
    \label{fig:chifermcontextr}
\end{figure}

As an example, in Fig.~\ref{fig:chifermcontextr} we present the topological susceptibility $\chi^{1/4}$ as a function of the squared lattice spacing $a^2$ for several temperatures, obtained from the interpolation of our data. We also present the results of the simple $a^2$ continuum extrapolation by bands of the corresponding color and in legend we show the chi-squared $\chi^2$ of these fits, what again confirms the relative robustness of the continuum extrapolation for $T\lesssim250$~MeV  and $T\gtrsim500$~MeV and difficulty of the continuum extrapolation in the intermediate region $250\mathrm{~MeV}\lesssim T \lesssim 500\mathrm{~MeV}$. Possibly this behaviour is attributed, up to 300 MeV,  to the sharp crossover, or even another transition in QCD at temperatures $T\sim 300$~MeV (see detailed discussion of this threshold later on). At higher temperatures around $400-500$~MeV the interpolation procedure may be complicated by possible outliers, and again the continuum extrapolation has large errors. Given this peculiarity of the data, in Fig.~\ref{fig:chiferm} the continuum extrapolated curve has large uncertainties. In order to estimate systematic error we performed continuum extrapolation using all three ensembles B64, C80 and D96 and only two finest  ensembles C80 and D96 and take half-sum as our final value and half-difference between two extrapolations as systematic uncertainty, which is included in the final error in Fig.~\ref{fig:chiferm}. We would like to note, that given that we need to interpolate our data between different points and large fluctuations of the data between different ensembles in the intermediate temperature region, it is very difficult to perform the continuum extrapolation, however we believe that this procedure allows us to get a reasonable estimate of the continuum extrapolated data and its uncertainty.

\subsection{The region around criticality}

In the lowest temperature regime  $T \simeq T_c$  the lattice effects are moderate,
a good continuum extrapolation is available 
(see again Fig.~\ref{fig:chiferm}). Note that the estimate of the topological susceptibility we used relies on the restoration of chiral symmetry~--- this good agreement shows that the residual breaking  is small in the range of temperatures we have explored and further validates the fermionic method. 

\subsection{The crossover region}

 At slightly higher temperatures $T\gtrsim250$~MeV we observe
a faster decrease of the topological susceptibility. In this region, as discussed above, 
the continuum limit is affected by large uncertainties: they come from different sources. First, the results are noisier; second, the interpolation is more
difficult due to the fast change of the observables. Different interpolators give
consistent results, with an error band which is in general as large as the largest error on the interpolated points. Even if the continuum limit is noisier, different
lattice ensembles help building a consistent picture: 
the signals of a fast crossover are indeed visible in all different ensembles and by inspecting Fig.~\ref{fig:chiferm} again we see a consistently large slope 
even within the large errors. 
As an aside, the associated inflection points~--- estimated via logistic fits on
a restricted sets of points (shown in Fig.~\ref{fig:chiferm} as dashed dotted lines)~--- are 193, 253, 270 MeV for the different ensembles.
We also note that  our  older results \cite{Burger:2018fvb} with larger quark masses do show  some crossover, which apparently is  more pronounced at smaller  masses.

It is then worth noticing that this fast crossover
happens in the same region where others have observed phenomena suggestive of a phase transition \cite{Mickley:2024vkm,Glozman:2024ded,Aarts:2023vsf,Kotov:2021rah,Cardinali:2021mfh,Alexandru:2019gdm,Rohrhofer:2019qwq}.
When considering that the topological susceptibility is linked to the axial symmetry,
it becomes natural to associate the crossover with the onset of the
region with trivial topology, as described by DIGA. 

\subsection{DIGA regime}
\label{sec:diga}

At even larger temperatures, up to 600 MeV, we can study the approach to the DIGA regime. We have done this either by extrapolating to the continuum limit the results of
fits on different ensembles, as well as by fitting directly the continuum results
described above.

\begin{figure}
\centering
\includegraphics[width=0.45\linewidth]{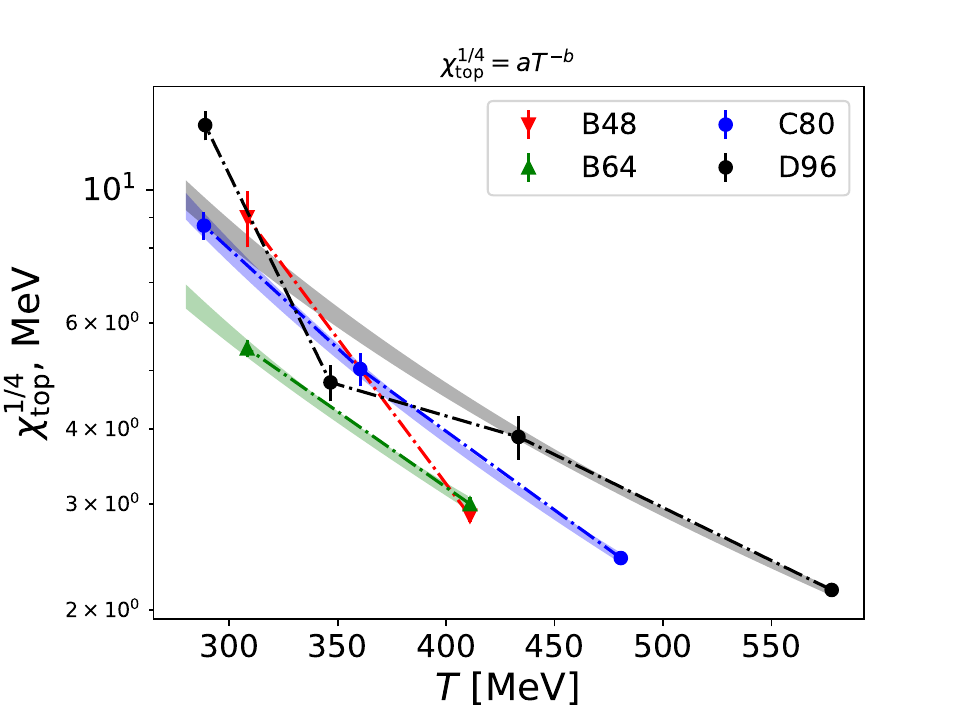}
\includegraphics[width=0.45\linewidth]{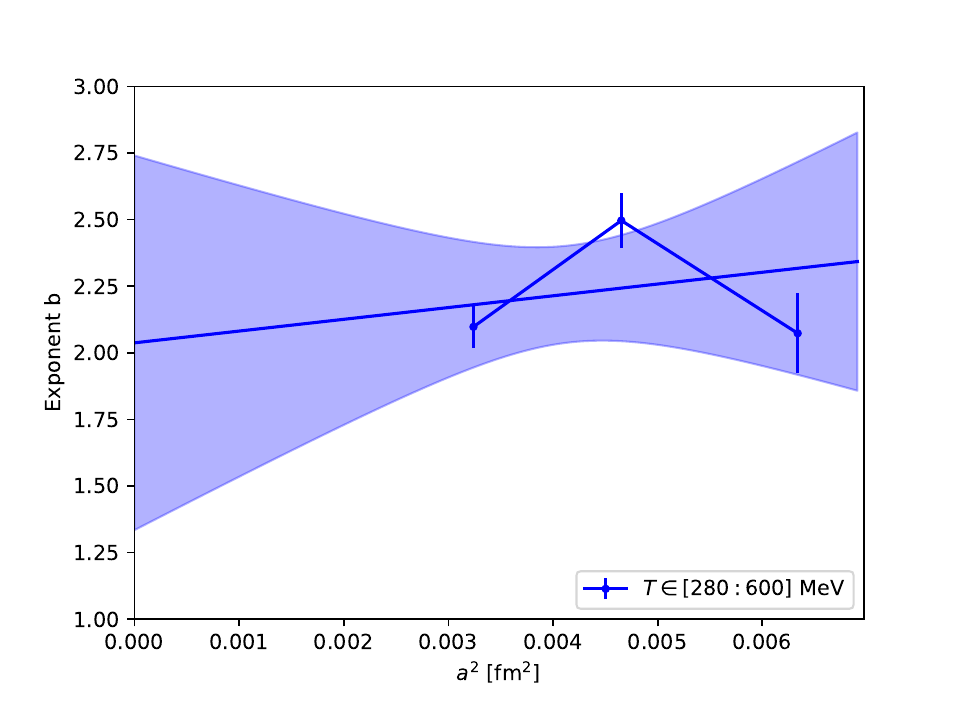}
\caption{Fit of the topological susceptibility to the DIGA inspired form: $\chi_{\mathrm{top}}^{1/4}=AT^{-b}$ for different ensembles in the temperature range $T\in [280,600]$~MeV (left panel). Continuum extrapolation of the exponent $b$ from this fit (right panel).}
\label{fig:digaexp}
\end{figure}

\begin{figure}
\centering
\includegraphics[width=0.7\linewidth]{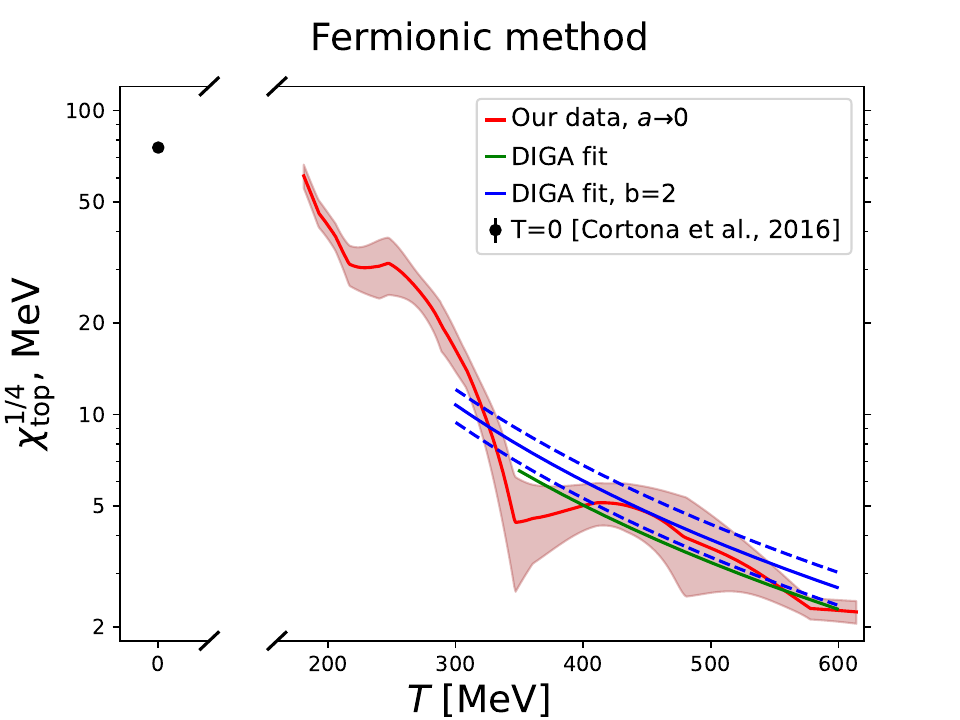}
\caption{A summary plot for the DIGA comparison: we show the continuum extrapolated band from fits on individual ensembles (blue lines), which compares well with the continuum results (red lines). We also show the fit with open parameters to the continuum results with a green line, again
in good agreement with the DIGA form. For comparison we present the topological susceptibility at zero temperature $T=0$ from  \cite{GrillidiCortona:2015jxo}. }
\label{fig:finalferm}
\end{figure}

For the behaviour on individual ensembles,   fits with an open exponent
extrapolates to a continuum value consistent either with $ b = (1.92, 2)$, corresponding to $N_l = 2,3$  
respectively, Fig. \ref{fig:digaexp}.
Alternatively, in the Fig.~\ref{fig:comparediga} 
we have shown fits with a  DIGA exponent constrained to the $N_l=3$ value.  In this region the continuum extrapolation gives reasonable results, as we have discussed. 
In Fig.~\ref{fig:finalferm} we show these fits, together  with  the extrapolated band.

Finally, we can fit directly the extrapolated results. This procedure has to be done with some care, since the extrapolated results were obtained from the interpolation of raw data at fixed temperature values. For this reason, we take the fit range to be $[350:600]$~MeV with a slightly higher value of the starting temperature. This temperature value approximately corresponds to the lowest raw data temperature points for C and D ensembles above the crossover behaviour $T\approx300$~MeV and thus is not contaminated during the interpolation procedure by the data from temperatures below 300~MeV. In the temperature range $[350:600]$~MeV we took temperature points with steps $40$ MeV and fit the topological susceptibility in these points. Since this temperature step roughly corresponds to the steps between different ensembles (see Fig.~\ref{fig:chiferm}), we  assumed that extrapolated data for different temperatures are statistically independent and performed an uncorrelated fit. The results of this fit are also presented in Fig.~\ref{fig:chifermcontextr} by a green line and they agree with the continuum extrapolation of the constrained fits on individual ensembles with $\sim\sigma$ difference. The fitted value of the exponent $b$ is $b=1.95\pm0.37$, which is compatible with both $b$ values $b=1.92,\,2$. 

The situation is summarised in Fig.~\ref{fig:finalferm}. We conclude that the results for the exponent are consistent with a dilute instanton gas above 300 MeV. As noted by others, the amplitude~$A$, see Eq.~\eqref{eq:diga}, is not matching the predicted DIGA value, even taking into account the recent improvements~\cite{Boccaletti:2020mxu}. However, this discrepancy may well be attributed to the limitation of the analytic approach, and the consistency with the exponent is a sufficient indication of the DIGA regime.  It is interesting, that the results from the gluonic method reported in the next section also suggest the onset of DIGA behaviour at $\sim 300$~MeV.

 \subsection{Comparison with other results}

 \begin{figure}
     \centering
     \includegraphics[width=0.7\linewidth]{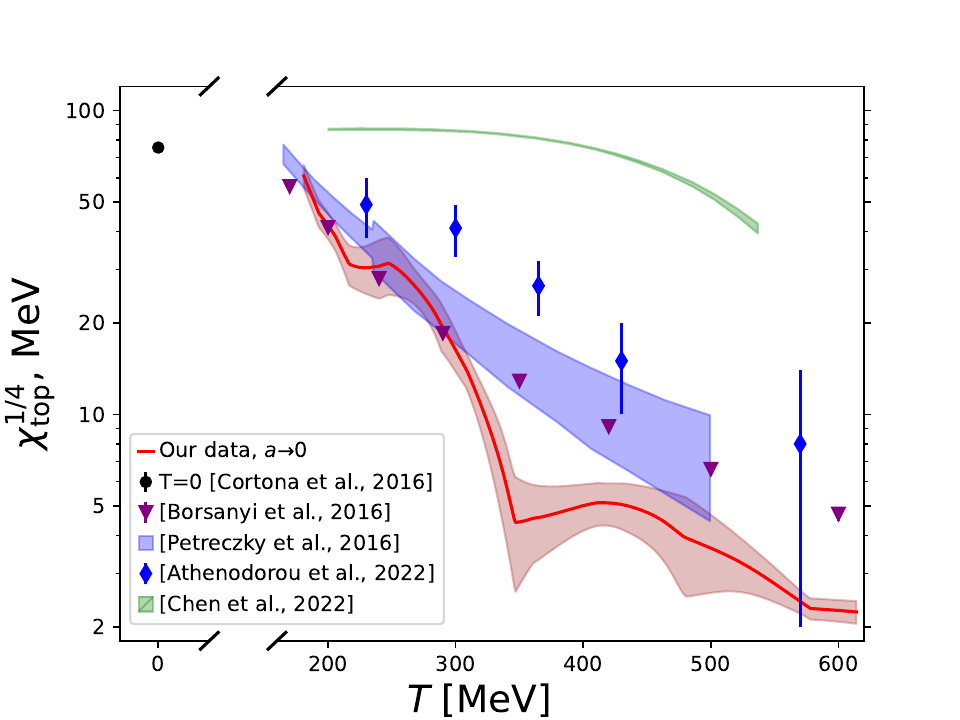}
     \caption{Comparison of our results and results of other groups \cite{Borsanyi:2016ksw,Petreczky:2016vrs,Athenodorou:2022aay,Chen:2022fid}. For comparison we present the topological susceptibility at zero temperature $T=0$ from  \cite{GrillidiCortona:2015jxo}.}
     \label{fig:chifermcomparison}
 \end{figure}

In Fig.~\ref{fig:chifermcomparison} we present our results for the topological susceptibility from the fermionic method and results obtained by other groups for physical or near physical pion mass. A more complete compilation with older
results is available in \cite{Aarts:2023vsf}. It is clear now that the consensus between different groups is not reached yet, our results seem to be in general on the lower side. It is interesting to note that the results from WB~\cite{Borsanyi:2016ksw} and from HotQCD\footnote{Note that in~\cite{Petreczky:2016vrs} the pion mass was slightly higher than physical, $m_{\pi}=160$~MeV.}~\cite{Petreczky:2016vrs} are consistent with ours at low temperatures, but are larger than ours around the crossover and above $T\sim 250$~MeV. The results of \cite{Athenodorou:2022aay} are slightly higher than data in \cite{Borsanyi:2016ksw,Petreczky:2016vrs} and our results. Finally, in \cite{Chen:2022fid} even higher values for the topological susceptibility are obtained. As for the transition to the DIGA behaviour, in~\cite{Borsanyi:2016ksw} the crossover is not visible, although in~\cite{Petreczky:2016vrs} the authors argue about the onset of DIGA behaviour around $T\sim 250$~MeV. A crossover to DIGA is apparent in the recent paper \cite{Gavai:2024mcj}. This paper  analyses the disconnected susceptibility and finds its peak close to the pseudo-critical temperature, and confirms
that for $T=186$~MeV the anomaly is still broken. 
In summary, there are remaining quantitative differences among methods.
The approach to DIGA is consistently observed as we will also discuss in the gluonic Section. 

More interestingly, apparently the fast crossover to DIGA is more apparent in the fermionic method~--- one may speculate that this is related to some unappreciated behaviour of the
non-singlet residual contribution~--- however at the moment there is no evidence for this.

\begin{figure}
    \centering
    \includegraphics[width=0.45\linewidth]{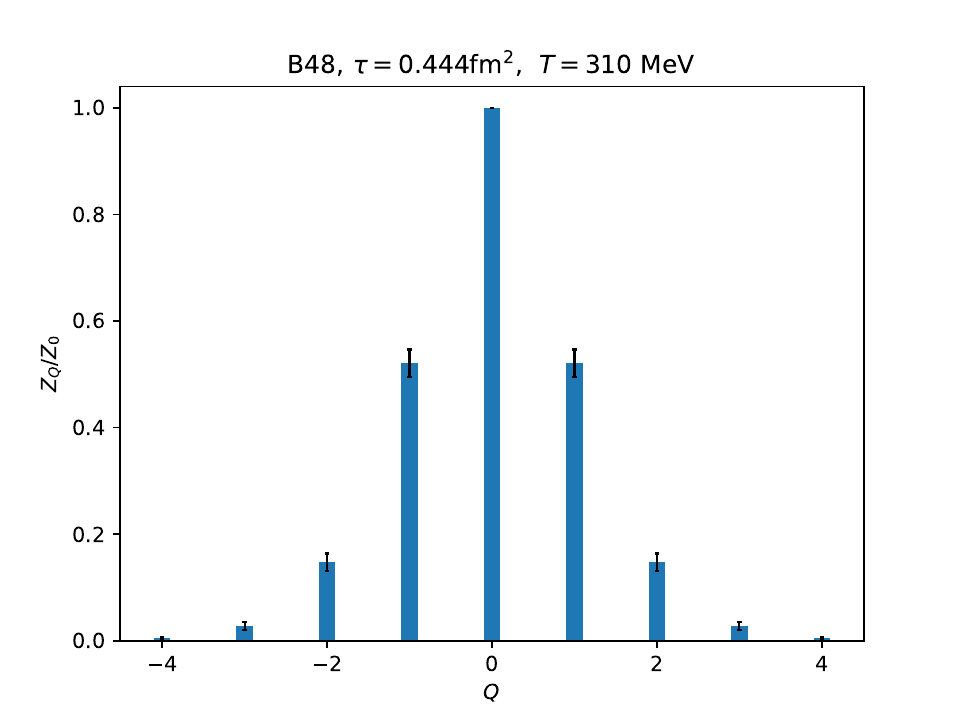}
     \includegraphics[width=0.45\linewidth]{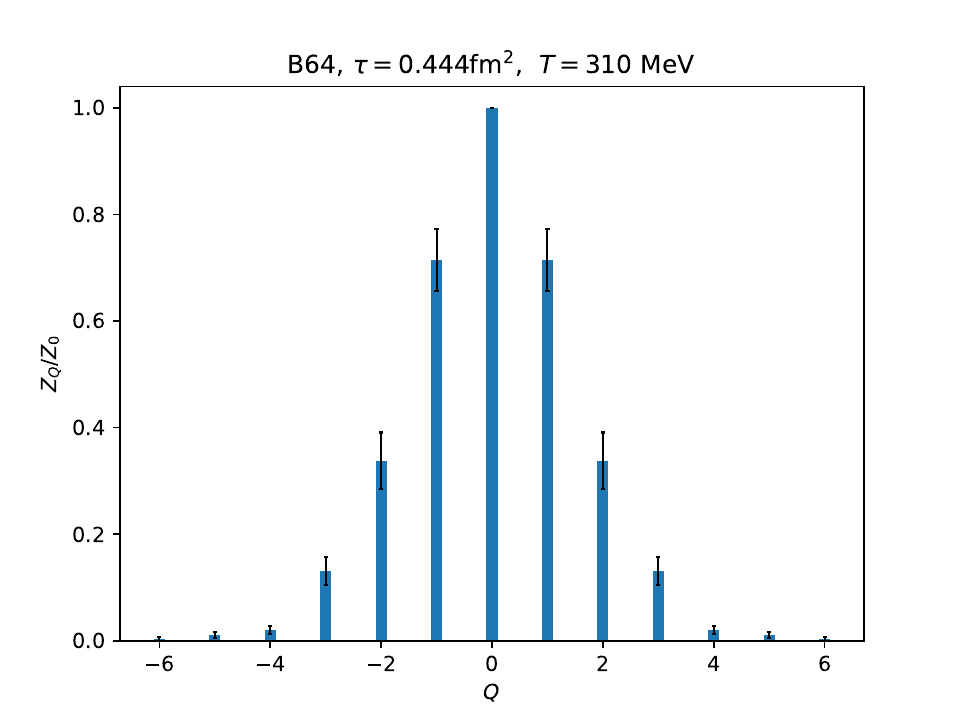}
    \includegraphics[width=0.45\linewidth]{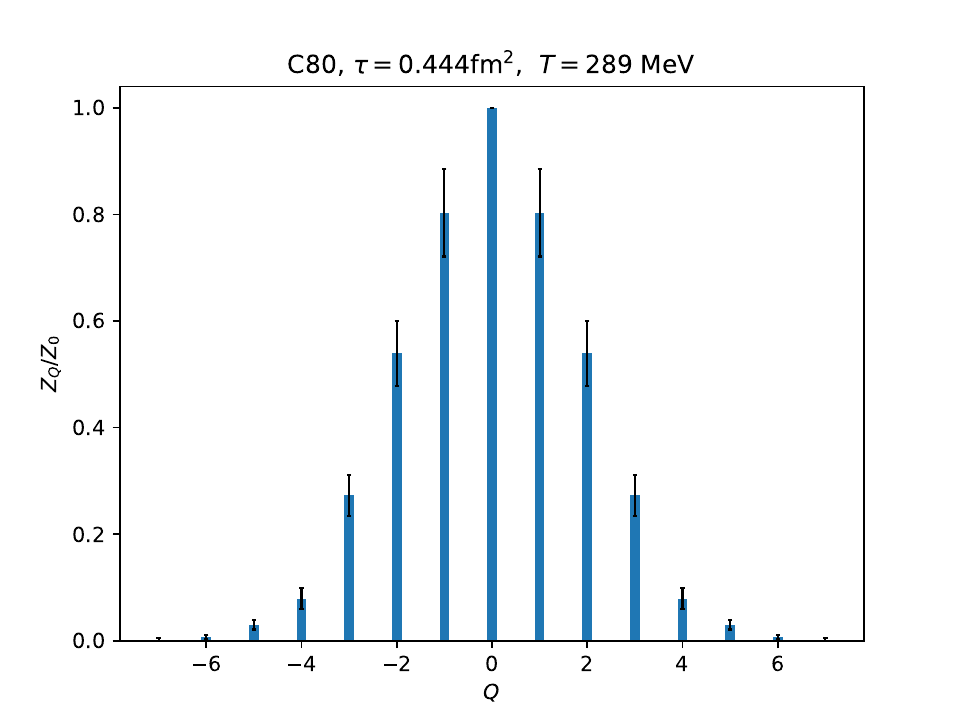}
    \includegraphics[width=0.45\linewidth]{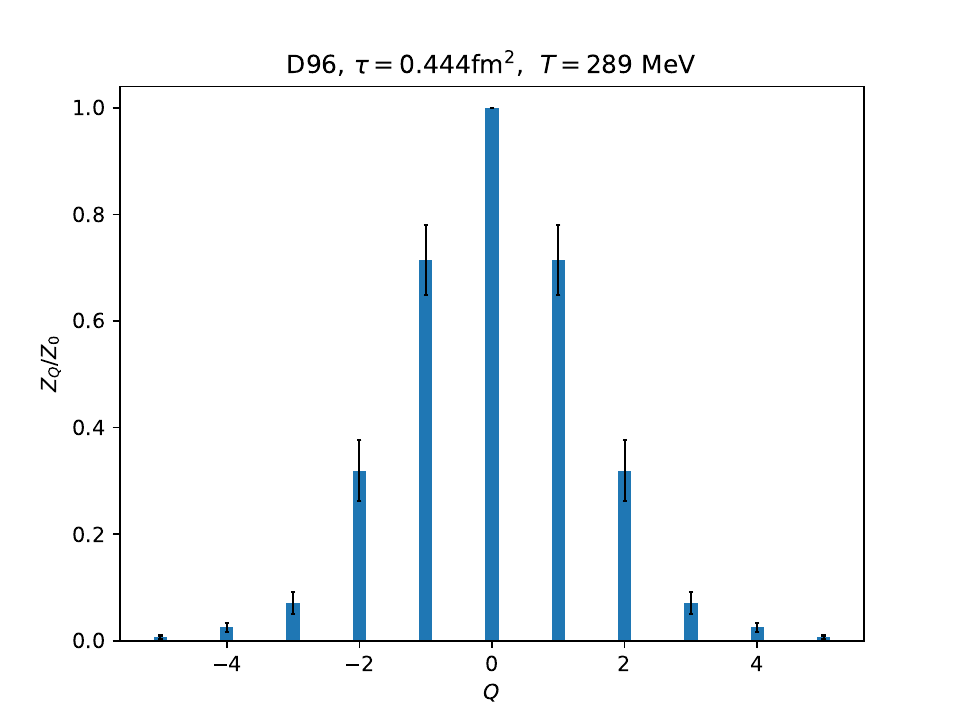}
    \caption{Distribution of the topological charge $Q$ on all four ensembles for temperature $T\approx310$~MeV.}
    \label{fig:histo}
\end{figure}

\section{Results on topology from gluonic method}
\label{sec:resultsgluonic}

In this section we discuss the results for the topological susceptibility and related observables from gluonic (field theoretical) definition of the topological charge given by Eq.~(\ref{eq:qgluonicdef}). Contrary to fermionic definition, field theoretical discretization of the topological susceptibility gives the value of the topological charge $Q$ on each configuration, what allows us to study the distribution of the topological charge, topological susceptibility and higher order cumulants of the distribution, e.g. $b_2$, as well as to
experiment with the reconstruction of the Free Energy.

\subsection{Topological charge histograms, cooling, evolution of the distribution}

In Fig.~\ref{fig:histo} we present examples of the distribution for the topological charge $Q$ for one relatively high temperature $T\approx300$~MeV.
We present the histograms for all four ensembles for one close temperature value. Note that since we are doing simulations in the fixed scale approach, our values of temperatures for each ensemble are discrete and we cannot take exactly the same values of temperature for all ensembles. For these plots we take the gradient flow time $\tau=0.444$~fm$^2$. 

\subsection{Topological susceptibility}

\begin{figure}
    \centering
    \includegraphics[width=0.45\linewidth]{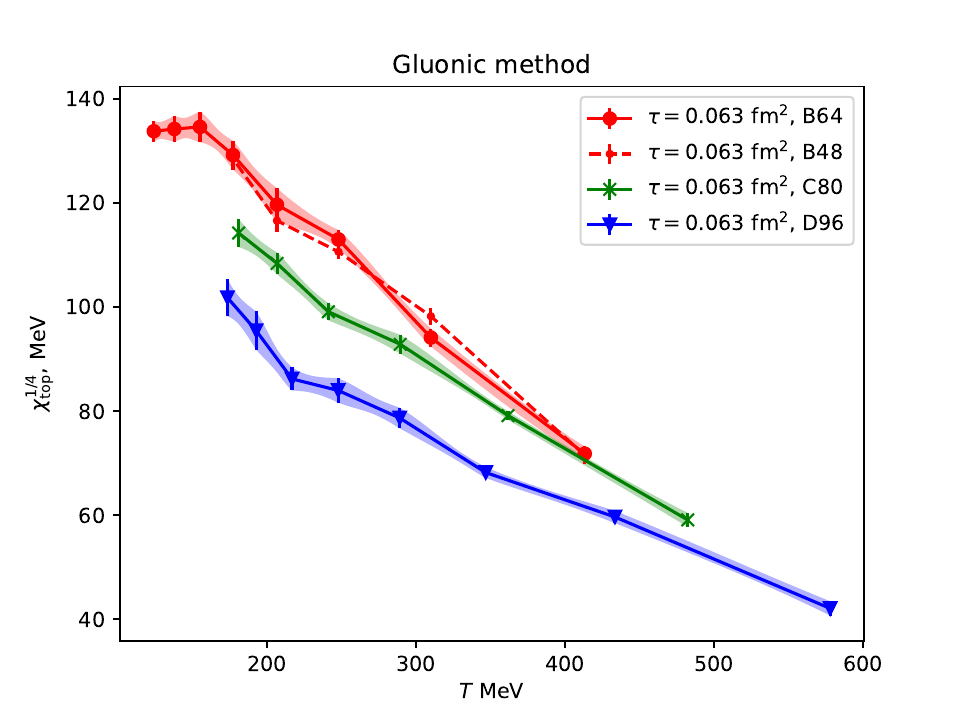}
    \includegraphics[width=0.45\linewidth]{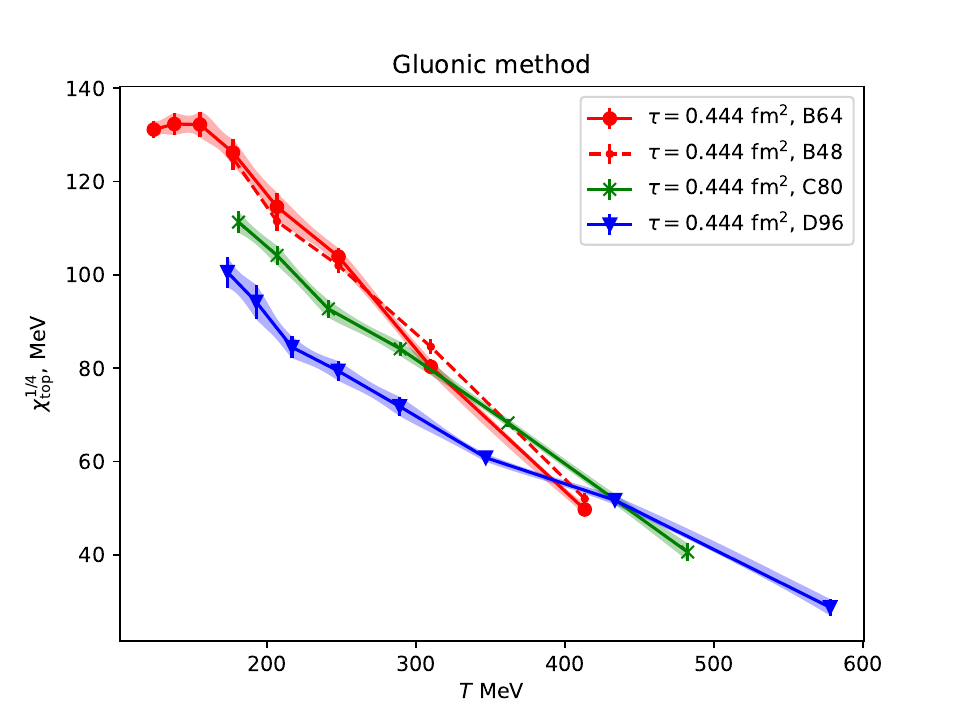}
    \caption{Temperature dependence of the topological susceptibility $\chi^{1/4}$ for all four ensembles and two gradient flow times $\tau=0.063$ fm$^2$(left plot) and $\tau=0.444$ fm$^2$(right plot).}
    \label{fig:topsusfixedflow}
\end{figure}

In Fig.~\ref{fig:topsusfixedflow} we present the temperature dependence of the topological susceptibility $\chi^{1/4}$ for all four ensembles and two gradient flow times $\tau=0.063$~fm$^2$ and $\tau=0.444$~fm$^2$. These data were obtained after the rounding procedure. First, by comparing data for ensembles B64 and B48, shown by red solid and and red dashed lines, we can conclude that finite volume effect are small (within errorbars) and we neglect them in our discussion. Note that we do not present data for $N_t=4$ here. The reason for it is because for high temperatures the topological susceptibility is small and our simulations for the current volumes are mainly in the zero sector $Q=0$, preventing a reliable measurement of the topological susceptibility for these points. In order to get data between the temperature values where we performed the simulations, we performed a linear and monotonic cubic spline interpolation of our data in the same way as we did within the fermionic method. The results of the interpolating procedure are presented in Fig.~\ref{fig:topsusfixedflow} by the corresponding color band. Using the results of this interpolator we can get the values of the topological susceptibility for all three ensembles at one temperature and perform a continuum extrapolation. 

By comparing the data for ensembles B64, C80 and D96, we clearly see very large cutoff effects, requiring a continuum extrapolation of our data. To perform a continuum extrapolation we used a second-order polynomial function in $a^2$:

\begin{equation}
    \chi^{1/4}(a^2)=\chi^{1/4}(a^2=0)+A_2a^2+A_4a^4
    \label{eq:contextr}
\end{equation}

Note that lattice artifacts and correspondingly coefficients $A_2$ and $A_4$, in principle, could depend on the flow time $\tau$, however the continuum extrapolated value $\chi^{1/4}(a^2=0)$ should be flow time independent. For this reason we performed a joint fit of our data at two flow times with a function in Eq.~(\ref{eq:contextr}):

\begin{equation}
    \chi^{1/4}(a^2,\tau=\tau_{1,2})=\chi^{1/4}(a^2=0)+A_2(\tau=\tau_{1,2})a^2+A_4(\tau=\tau_{1,2})a^4.
\end{equation}

\begin{figure}
    \centering
    \includegraphics[width=0.6\linewidth]{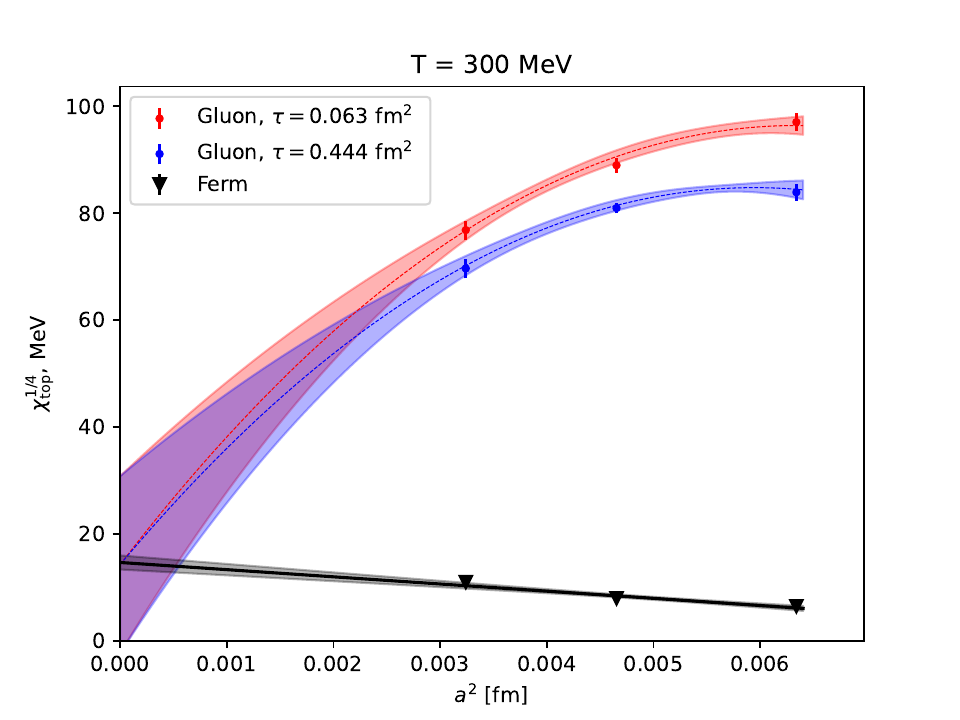}
    \caption{Continuum extrapolation of the topological susceptibility $\chi^{1/4}$ for temperature $T=300$~MeV. For the gluonic definition a joint fit for the data for two gradient flow times $\tau/a^2=0.063$ fm$^2$ and $0.444$ fm$^2$ is presented. Additionally we show the continuum extrapolation of the topological susceptibility from the fermionic definition.}
    \label{fig:contextr300}
\end{figure}

Since we interpolate our data, we can perform a continuum extrapolation for any temperature $T\in[185, 410]$~MeV. The largest possible temperature $T\approx410$ MeV  at which we can study the continuum extrapolation of the gluonic definition corresponds to $N_t=6$ for B64 ensemble.
In Fig.~\ref{fig:contextr300} we present the dependence of the topological susceptibility $\chi^{1/4}$ on the lattice spacing for both gradient flow times and a joint continuum extrapolation of these data for one temperature $T=300$~MeV. We also present the results of the topological susceptibility, determined from the fermionic method with its continuum extrapolation. Again, one clearly sees large cutoff effects for the gluonic definition of the topological susceptibility and, consequently, large errors in the continuum extrapolated values of $\chi^{1/4}(a^2=0)$. However, quite remarkably, the results of the fermionic and gluonic definitions agree with each other.

\begin{figure}
    \centering
    \includegraphics[width=0.7\linewidth]{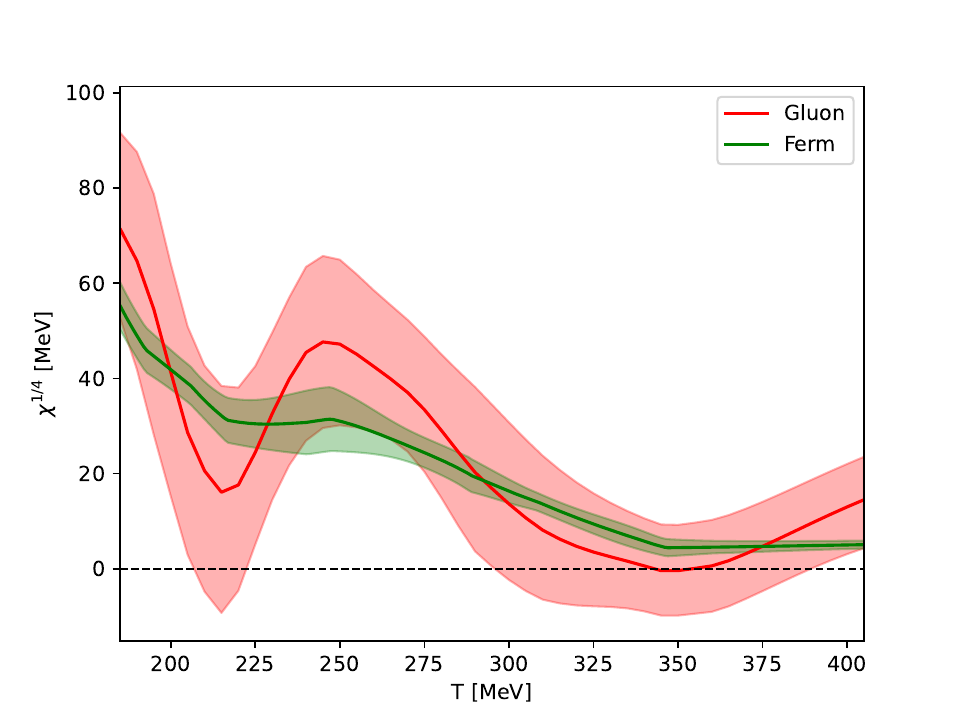}
    \caption{Continuum extrapolated topological susceptibility $\chi^{1/4}$ as a function of temperature from gluonic and fermionic approaches.
    }
    \label{fig:chifinal}
\end{figure}

In Fig.~\ref{fig:chifinal} we present the final temperature dependence of the continuum extrapolated topological susceptibility determined from both fermionic and gluonic definitions. We again see a clear agreement between both methods, although, as expected, gluonic definition has larger errors. This result strongly suggests that we have lattice artifacts under control and that we can use any of these methods. 

\subsection{Higher moments of the topological charge distribution}

\begin{figure}
    \centering
    \includegraphics[width=0.45\linewidth]{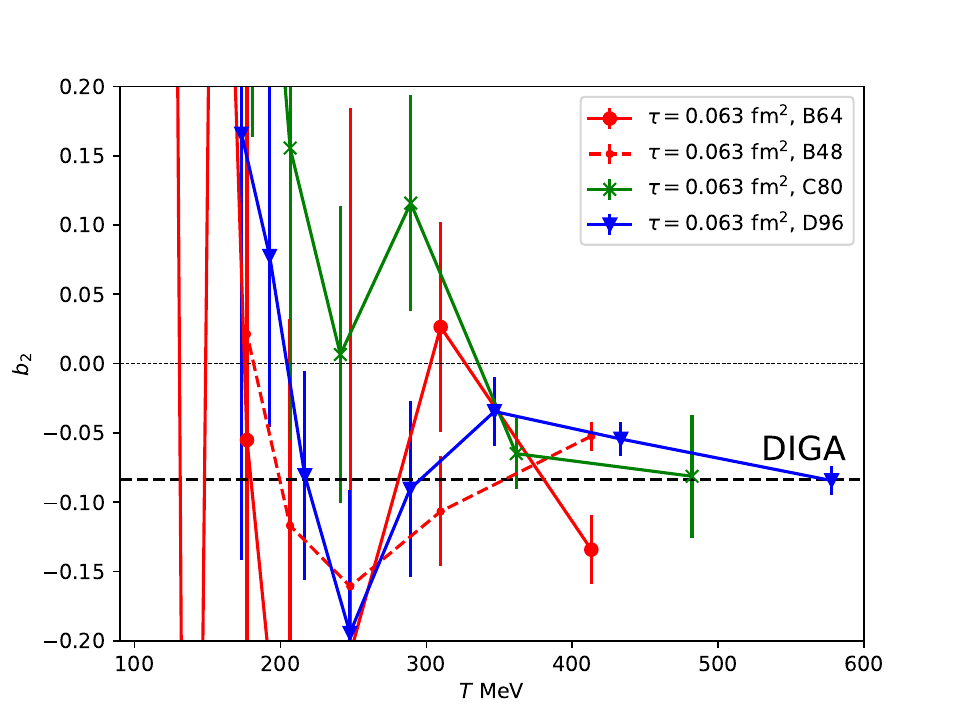}
    \includegraphics[width=0.45\linewidth]{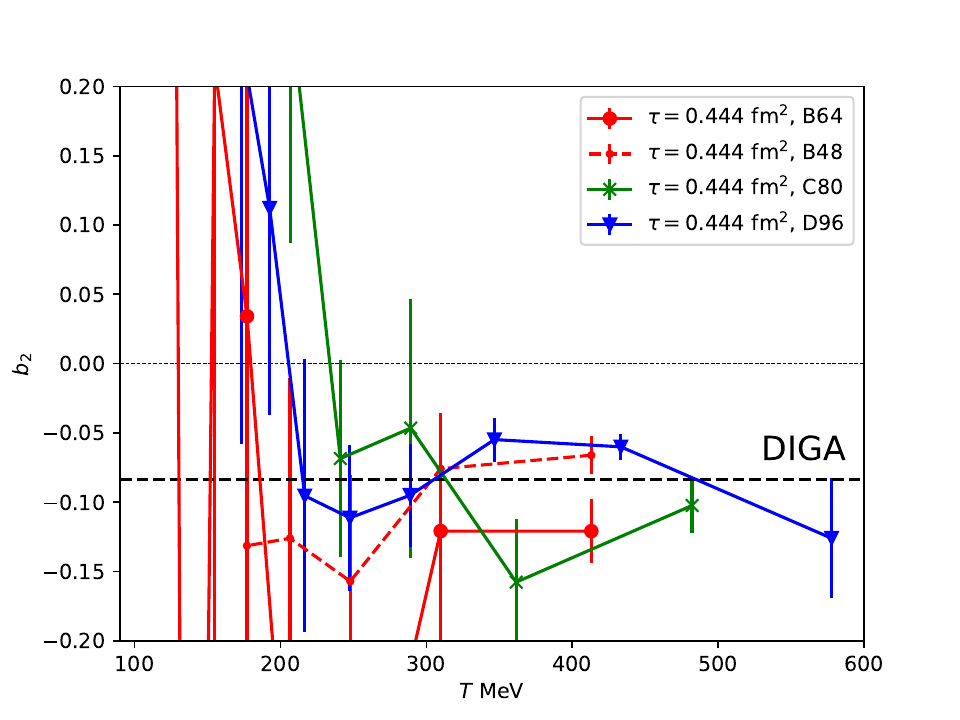}
    \caption{Temperature dependence of the kurtosis $b_2$ of the topological charge distribution for all ensembles at two gradient flow times $\tau=0.063$~fm$^2$ (left) and $\tau=0.444$~fm$^2$ (right). Horizontal dashed line corresponds to the Dilute Instanton Gas Approximation $b_2=-1/12$.}
    \label{fig:b2}
\end{figure}

Using the gluonic definition of the topological charge, we extracted the values of the kurtosis $b_2$ of the topological charge distribution:

\begin{equation}
    b_2=-\frac{\langle Q^4\rangle-3\langle Q^2\rangle^2}{12 \langle Q^2\rangle}.
    \label{eq:b2}
\end{equation}

\begin{figure}
    \centering
\includegraphics[width=0.7\linewidth]{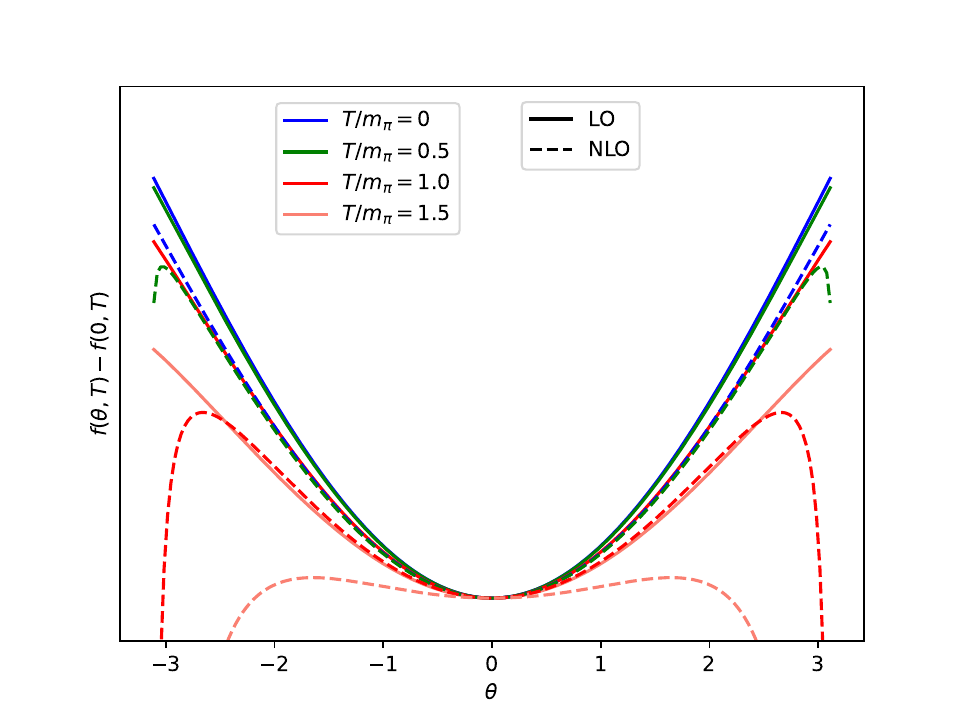}
    \caption{ChPT prediction for the Free Energy density $f(\theta,T)$. Different colors correspond to various temperatures as given in the legend. Solid and dashed lines represent LO and NLO approximations. Explicit expression are given in~\cite{GrillidiCortona:2015jxo}.}
    \label{fig:potential_chpt}
\end{figure}

In Fig.~\ref{fig:b2} we present the temperature dependence of the $b_2$ coefficient for all four ensembles and two gradient flow times $\tau=0.063$~fm$^2$ (left) and $\tau=0.444$~fm$^2$ (right). We also show the prediction of the DIGA behaviour $b_2=-1/12$ by a dashed horizontal line. The data are quite noisy and we cannot perform a controlled continuum extrapolation. However, it can be clearly seen that for all ensembles the data become less noisy at high temperature and start converging to the DIGA behaviour at $T\gtrsim 300$ MeV, threshold of the DIGA behaviour determined from fermionic definition. Again, as with the topological susceptibility, we can conclude that gluonic definition of the topological charge and its distribution is consistent with the predictions, obtained by fermionic definition (namely, onset of DIGA behaviour at $T\sim$ 300 MeV), albeit much noisier.

\subsection {Towards the Grand Canonical Partition Function}

Using the distribution of the topological charge $Z_Q$ we can calculate the partition function of QCD at nonzero $\theta$-angle $Z(\theta,T)$, and the associated Free Energy density $f(\theta,T)$: 
the  partition function  may be decomposed into a sum over partition functions $Z_Q$ each of which is associated with a fixed topological sector $Q$.
\begin{equation}
\begin{split}
    Z(\theta,T)=\sum_Q Z_Q(T)e^{iQ\theta},\\
    f(\theta,T)=-\frac{1}{V}\log\frac{Z(\theta,T)}{Z(0,T)}.
\end{split}    
\end{equation}

In our approach we will estimate $Z_Q$  from the distribution of the topological charge, see, e.g. Fig.~\ref{fig:histo}.  The reasonable agreement between fermionic
and gluonic estimates of the topological susceptibility gives us some confidence in  this procedure.

 The knowledge of the GCPF and of the Free Energy  at low and moderately high temperatures  so far comes from Chiral Perturbation Theory and Chiral Random Matrix Theory. The latter
 has been shown to be equivalent to QCD in the Leutwyler–Smilga regime \cite{Verbaarschot:2000dy}. At high temperatures we have the DIGA results.

 \begin{figure}
    \centering
\includegraphics[width=0.45\linewidth]{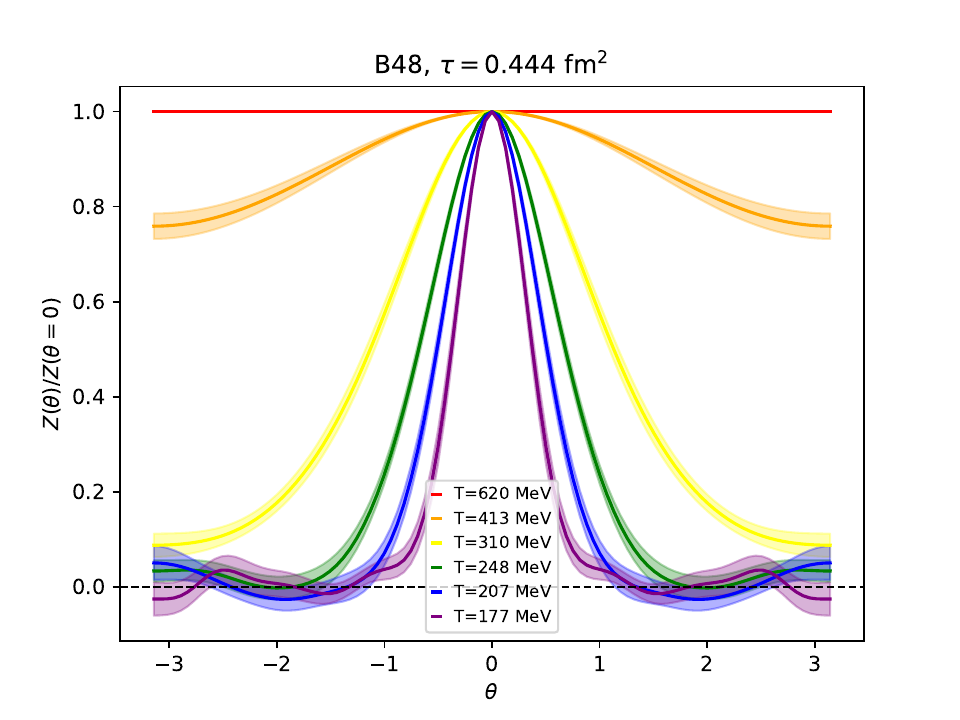}
\includegraphics[width=0.45\linewidth]{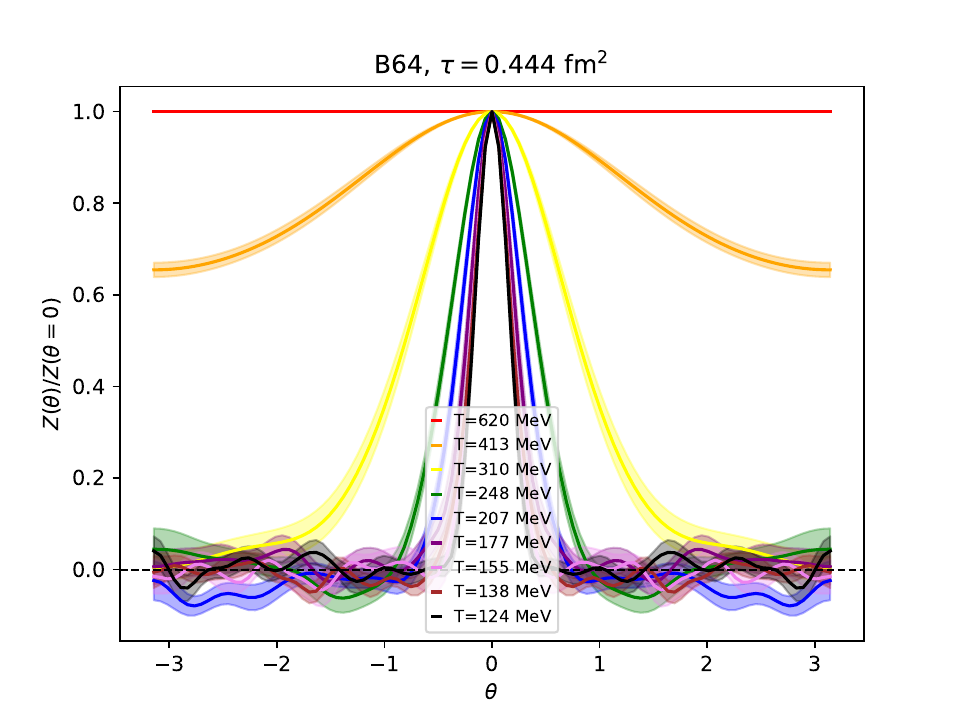}
\includegraphics[width=0.45\linewidth]{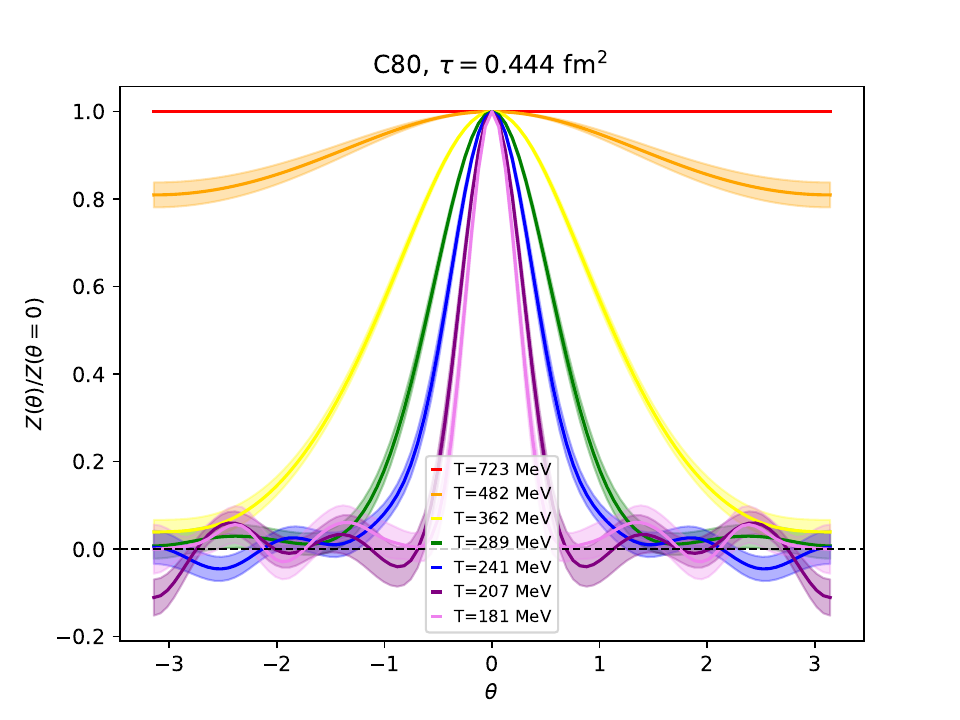}
\includegraphics[width=0.45\linewidth]{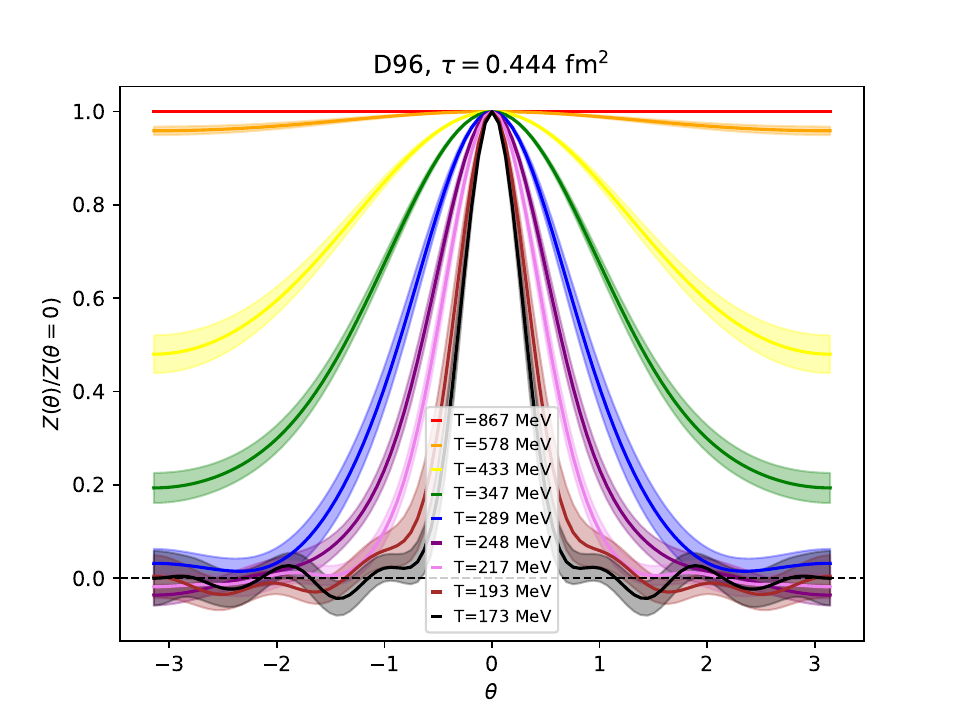}
    \caption{QCD partition function $Z(\theta)/Z(\theta=0)$ as a function of $\theta$-angle for different temperatures. Various plots correspond to various ensembles. Gradient flow time $\tau=0.444$~fm$^2$.}
    \label{fig:partitionfunction}
\end{figure}

At zero temperature Chiral Perturbation Theory predicts  the well know $\cos(\theta/2)$ dependence \cite{GrillidiCortona:2015jxo} for the energy. In the same paper, the temperature corrections have been estimated as well, see Fig.~\ref{fig:potential_chpt}. The main effect when temperature increases is the modification of the cusp at $\theta = \pi$:  the derivative around $\theta = \pi$ gets smaller. As the cusp is associated with a  discontinuity of the topological charge, the transition weakens with temperature.
It is well known, since the early work by Gasser and Leutwyler \cite{Gasser:1987ah}, that chiral perturbation theory breaks down at temperature of around $100$~MeV~--- this is natural from the perspective of an effective theory, where the symmetry pattern plays an important role. Anyway this interval suffices to appreciate the deformation of the energy, confirmed by the consistence of LO and NLO corrections in Fig.~\ref{fig:potential_chpt}. If, just as an exercise, we extend the results of ChPT at higher temperature we observe instabilities  at $T=300$~MeV. Similar phenomenon, manifested as negative values of the topological susceptbility, was also  observed in \cite{GomezNicola:2019myi}.

\begin{figure}
    \centering
\includegraphics[width=0.45\linewidth]{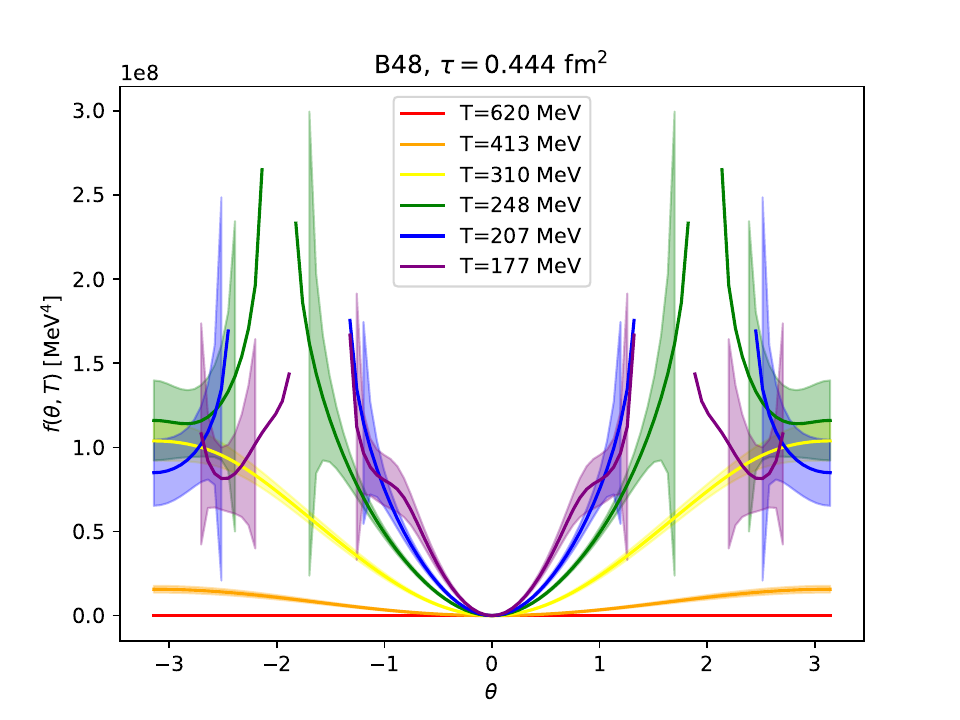}
\includegraphics[width=0.45\linewidth]{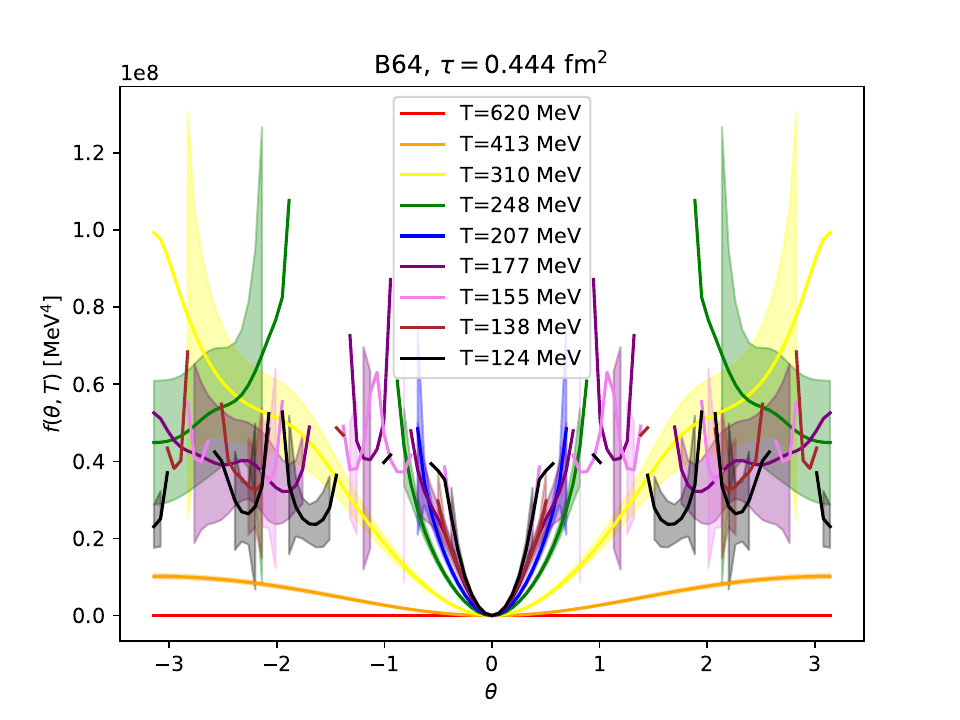}
\includegraphics[width=0.45\linewidth]{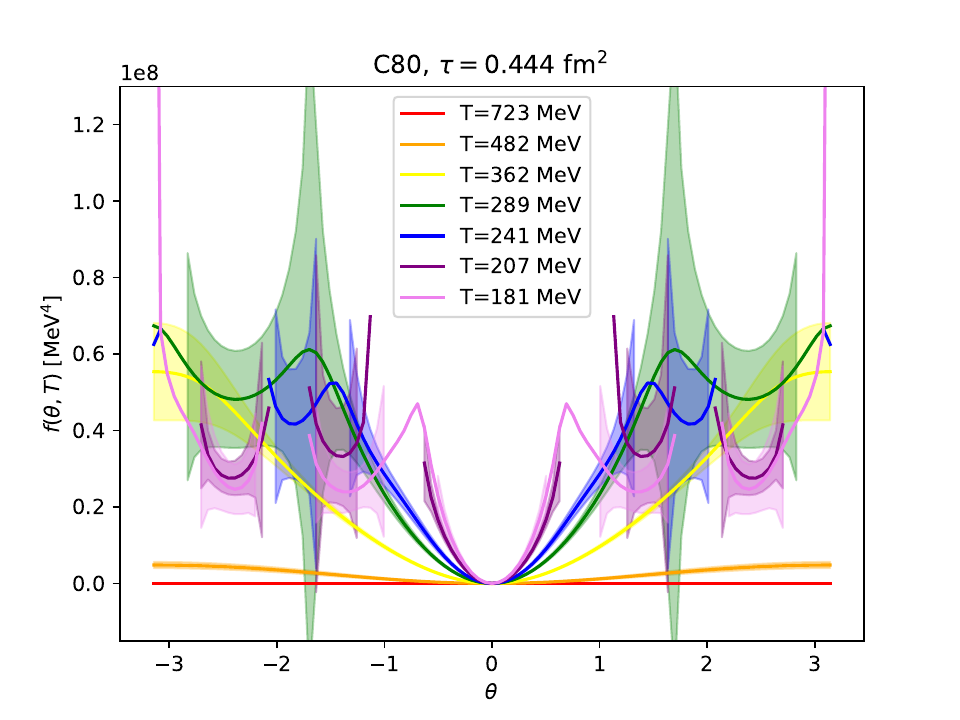}
\includegraphics[width=0.45\linewidth]{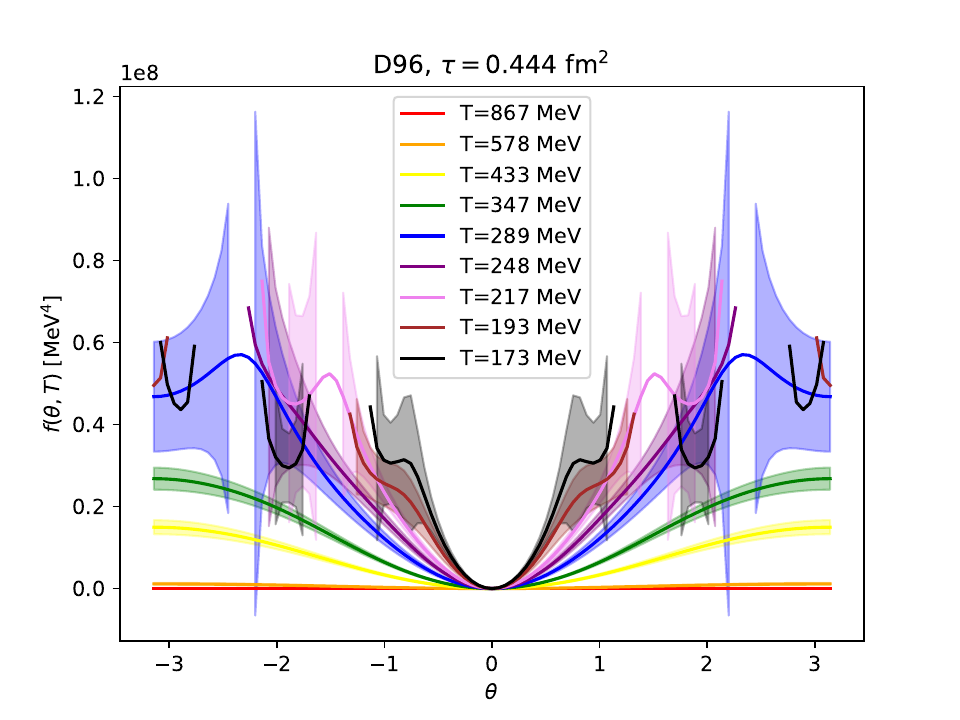}
    \caption{Free Energy density $f(\theta,T)$ as a function of $\theta$ for different temperatures, various panels correspond to four various ensembles. Gradient flow time $\tau=0.444$~fm$^2$. }
    \label{fig:potential}
\end{figure}

At very high temperature, as already mentioned, DIGA sets in, with the predicted behaviour
\begin{equation}
f(\theta, T)-f(\theta, 0) \simeq (1-\cos\theta)T^4 e^{-8\pi^2/g^2(T)},
\end{equation} 
The observed behaviour of the potential is then consistent with the approach to the analytic DIGA behaviour. While chiral perturbation theory predicts a weakening
of the cusp,  chiral perturbation theory alone does not help assessing the endpoint of
the cusp / first order phase transition in temperature. 

We note at this point that 
the seminal paper \cite{Smilga:1998dh} indicates a first order transition at 
$\theta = \pi$ for two degenerate light flavors. Further studies indicate the phenomenon is generic for all $n_f > 1 $~\cite{Creutz:2003xu}. The recent paper~\cite{DiVecchia:2017xpu} further analyzes the physics
around $\theta = \pi$. 

The properties around $\theta = \pi$ may be reflected in the behaviour of the zeros of the GCPF. 
This approach has been discussed in Ref. \cite{Akemann:2001ir}, 
and indeed there was observed an accumulation of zeros in the infinite volume limit in the complex $\theta$ plane
\begin{equation}
\lim_{V \to \infty} \theta  \equiv a  +i b \to \pi
\end{equation}
for $m \ne 0$, where $m$ is the quark mass.

\subsection{Results}

In short,  the issue of the $\theta$  and temperature dependence
of QCD may be  studied from (at least)  three interrelated points of view:

\begin{itemize}
    \item The analysis of the GCPF and the occurrence of zeros
    \item  The behaviour of the Free Energy as a function of $\theta$
    \item The occurrence of a first order transition at $\theta = \pi$
\end{itemize}

\begin{figure}
    \centering
\includegraphics[width=0.45\linewidth]{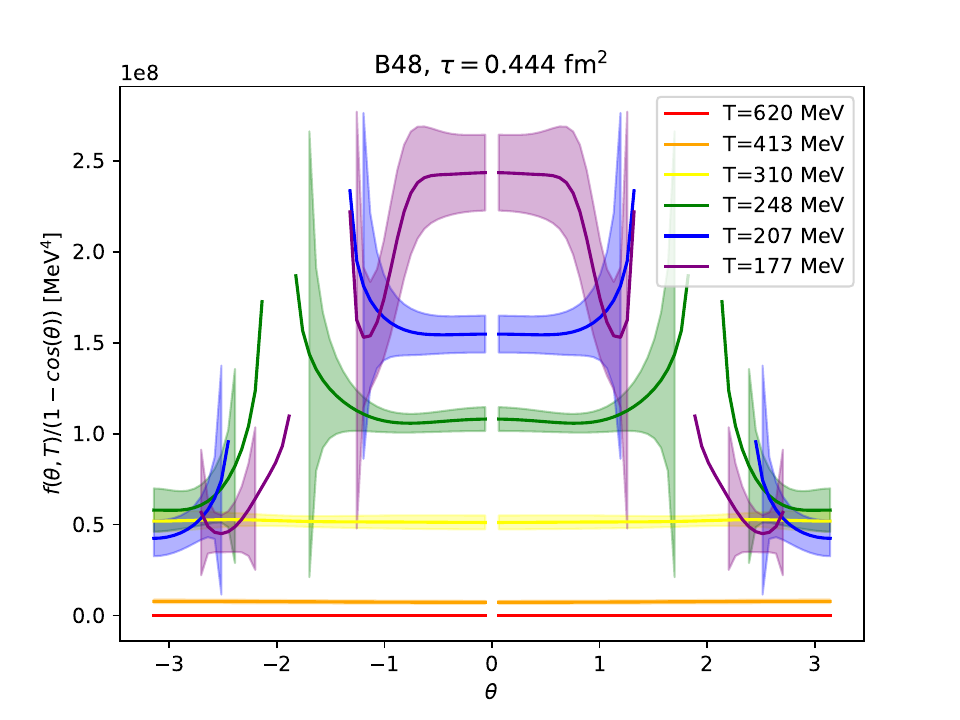}
\includegraphics[width=0.45\linewidth]{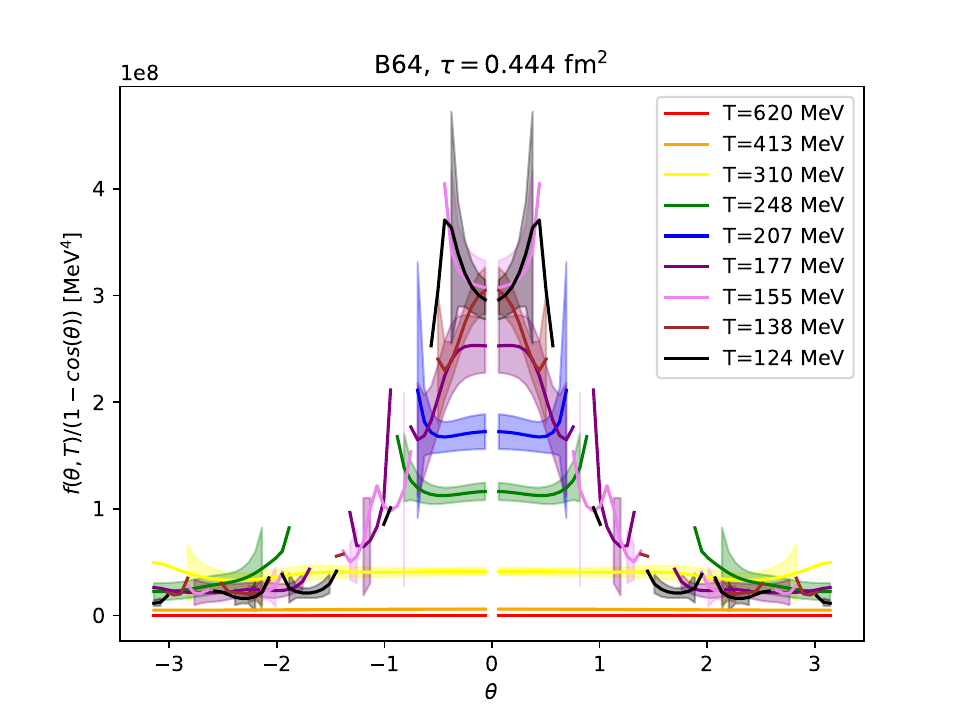}
\includegraphics[width=0.45\linewidth]{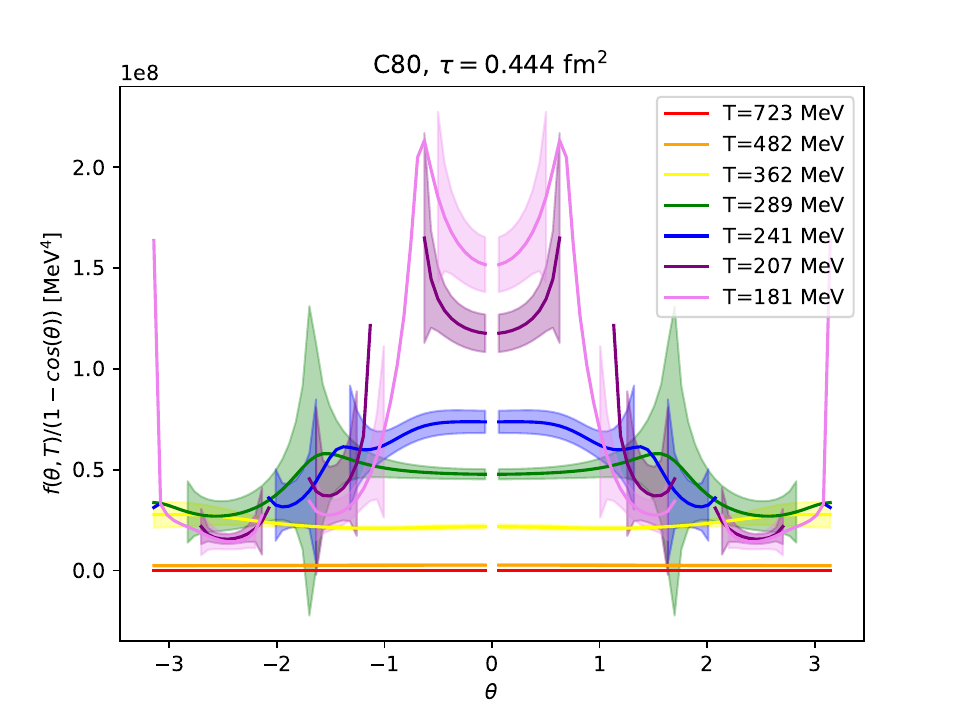}
\includegraphics[width=0.45\linewidth]{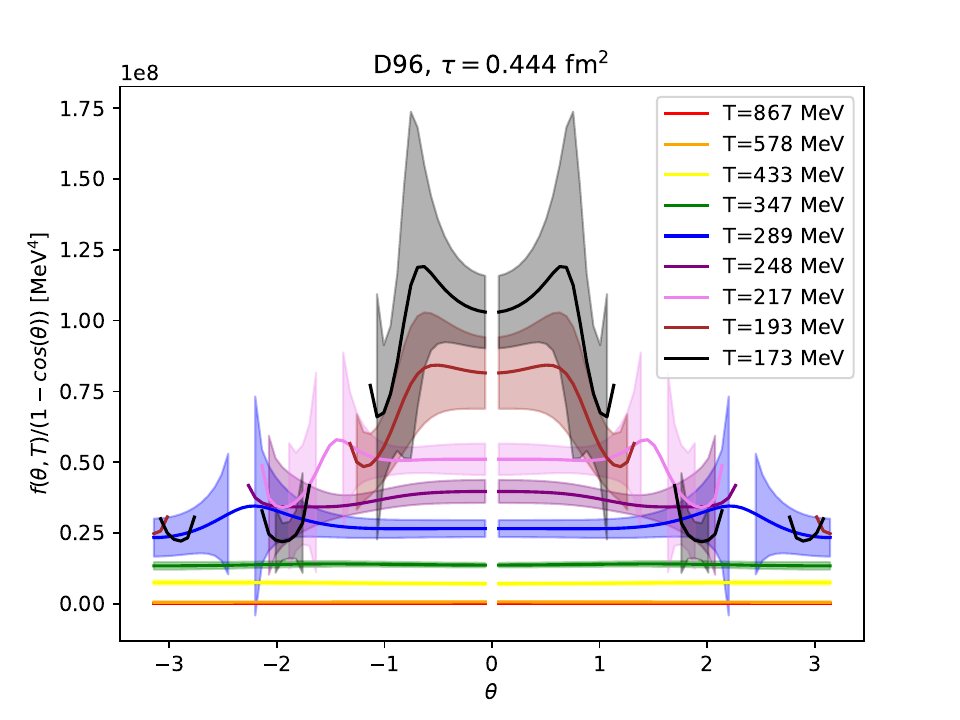}
    \caption{Free Energy density $f(\theta,T)$ divided by $1-\cos(\theta)$ as a function of $\theta$ for different temperatures, various panels correspond to four various ensembles. Gradient flow time $\tau=0.444$~fm$^2$. }
    \label{fig:potentialdivcos}
\end{figure}

In Fig.~\ref{fig:partitionfunction} we present the dependence of the QCD partition function $Z(\theta)/Z(\theta=0)$ at nonzero $\theta$-angle on the value of $\theta$ for all four ensembles under study, all temperatures. Gradient flow time is fixed $\tau=0.444$~fm$^2$. One can clearly note a very interesting behaviour: for all ensembles and high temperatures $T\gtrsim300$~MeV, the partition function is positive for all values of $\theta$. However, for small values of temperature below $300$~MeV the partition function drops down very quickly and becomes consistent with zero for sufficiently large nonzero values of $\theta$. In accordance with this, if one looks at the plot for the energy 
$f(\theta,T)$, presented in Fig.~\ref{fig:potential}, then for large temperatures $T\gtrsim300$~MeV it has an expected from DIGA $\cos$-like behaviour, while for lower temperature it blows at some $\theta$ value. In order to make this behaviour more clear, in Fig.~\ref{fig:potentialdivcos} we present the energy $f(\theta,T)$ divided by $1-\cos(\theta)$. We clearly see that at $T\gtrsim300$~MeV the ratio $f(\theta,T)/(1-\cos(\theta))$ is almost $\theta$-independent.

\begin{figure}
    \centering
    \includegraphics[width=0.45\linewidth]{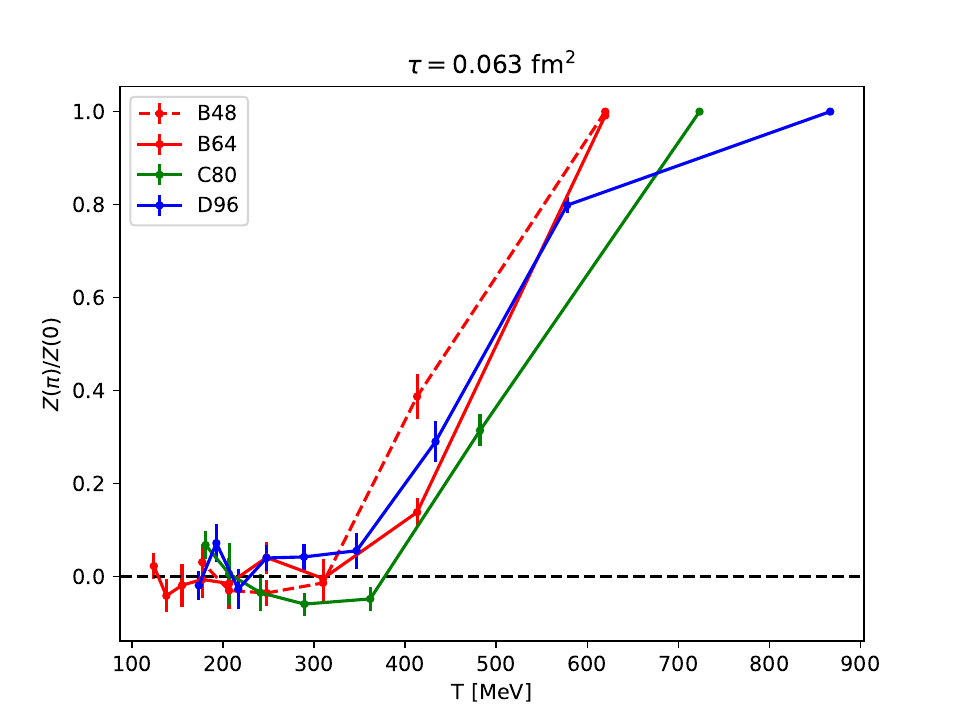}\includegraphics[width=0.45\linewidth]{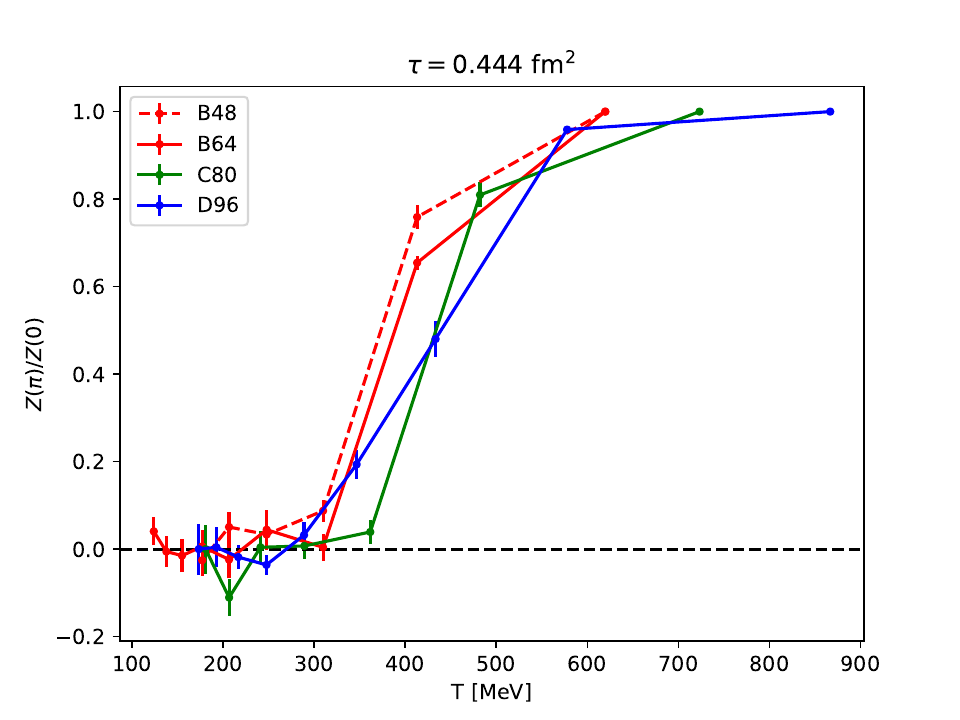}
    \caption{Ratio $Z(\pi)/Z(0)$ of the partition function $Z(\theta)$ at $\theta=\pi$ and $\theta=0$  as a function of temperature for all four studied ensembles. We present results obtained at two different GF times $\tau=0.063$ fm$^2$ (left panel) and $\tau=0.444$ fm$^2$ (right panel).}
    \label{fig:zpi}
\end{figure}

To monitor the behaviour of the zeros of the GCPF, we follow $Z(\theta = \pi)/Z(\theta = 0)$, see Fig.~\ref{fig:zpi}. 
What we observe is that $Z(\theta = \pi)/Z(\theta = 0)$ is 'very small' (perhaps compatible with zero) up to 
$T \simeq 300$~MeV. It would be very tempting to interpret the 'small' $Z(\pi)$
as a smoking gun for a nearby complex zero, evolving towards real zero in the infinite volume limit, as predicted by Ref.~\cite{Akemann:2001ir}. In this spirit, the observed $T=300$~MeV threshold may be interpreted 
as the threshold beyond which the Dashen phenomenon disappears -- but, admittedly, all this is a wishful thinking
as some guidance from analytic calculations would be desirable. At this stage,  we can
say that  our results indicate a significant change of the properties around $\theta = \pi$ at temperatures
of about $300$ MeV.

\section{Summary and conclusions}
\label{sec:conclusions}

We presented the results of our lattice study of the QCD topological properties at high temperatures $T>180$~MeV, carried out using $N_f=2+1+1$ Wilson twisted mass fermions at the maximal twist with  physical values of the quark masses.  We generated ensembles at three different lattice spacings and studied the continuum limit.

We presented the topological susceptibility determined from the fermionic method, using the disconnected chiral susceptibility,  and the gluonic method, using the  field-theoretical definition of the topological charge density, which was measured on the smeared configurations
with the Wilson flow. We confirm that the latter definition has larger cutoff effects, and, consequently, larger final uncertainty. Within this largish uncertainty,  we have found  agreement between the topological susceptibility determined from both gluonic and fermionic methods on our ensembles, see Fig.~\ref{fig:chifinal}. Note that the agreement between gluonic- and fermionic-based definitions of topological charge at high temperatures was also observed with staggered discretization in Ref.~\cite{Petreczky:2016vrs}  with a pion mass of $160$ MeV.  Similar results were obtained  using spectral projectors methods in Ref.~\cite{Athenodorou:2022aay}, again with staggered fermions.  Earlier studies with Wilson fermions have found agreement between gluonic and fermionic methods on a limited set  of temperatures and  larger quark masses \cite{Burger:2018fvb}. The present study has explored a large set of temperatures ranging from $180$~MeV to $600$~MeV at the physical point, using Wilson Fermions, with the inclusion of  dynamical strange and charm quarks. Earlier results with the same action were limited to larger pion masses~\cite{Burger:2018fvb}.

We used the gluonic definition to explore properties of the Grand Canonical Partition function at nonzero $\theta$ angle and the associated Free Energy. We have computed the kurtosis $b_2$, and reconstructed~--- to our knowledge for the first time~--- the GCPF $Z(\theta)$ and the Free Energy themselves. 

Our results  suggest three different regimes: low-temperature with $T<250$~MeV, fast and narrow crossover around $T\sim 300$~MeV and consistent with Dilute Instanton Gas Approximation behaviour above $T\gtrsim300$~MeV.  DIGA behaviour has been consistently observed in the temperature dependence of the topological susceptibility and of the kurtosis~$b_2$. Moreover, the exploratory calculation of the Free Energy afforded for the first time the direct observation of a $\cos(\theta)$ behaviour in the same range of temperature $T > 300$~MeV. The fast and narrow crossover~--- seen in the  topological susceptibility, in $b_2$, in the Free Energy $f(\theta,T)$ and the partition function $Z(\pi)/Z(0)$~--- happens in the same region where our previous study \cite{Kotov:2021rah}  suggests the limit of the scaling window for the transition. 
Other studies have associated this threshold to deconfinement, or to new pattern of symmetries.   

From the viewpoint of the phase diagram in the ($\theta, T$) plane, the observed near-zero value of $Z(\pi)$ at a temperature $T$ may be suggestive of  a nearby complex zero of  $Z$.  Effective models \cite{Akemann:2001ir} computed explicitly the zeros of $Z$ in the complex plane at zero temperature, and discuss their relation with spontaneous CP violation and the occurrence of
a first order phase transition at $\theta = \pi$. Then, it may be tempting to associate the growth of $Z(\pi)/Z(0)$ at $T \simeq 300$~MeV with the endpoint of this first order transition line, although of course further work, and in particular a finite size scaling analysis, would be needed to put this statement on firm grounds. 

In short summary, our  results build further confidence in the consistency of the fermionic and gluonic methods for the measurement of the topological susceptibility at $T > 300$~MeV, when applied to the same set of configurations. The results
 confirm the existence of a threshold in the QGP at temperatures around $300$~MeV.  We have presented a first
computation of the Grand Canonical Partition Function. That allowed the direct observation of the cos-like behaviour
of the potential at high temperatures, and of the  change of the behaviour at $\theta \simeq \pi$.

Concerning open problems, from a numerical viewpoint there remain clear
discrepancies among different groups at temperatures around $T \simeq 300$~MeV and above, where \cite{Borsanyi:2016ksw, Petreczky:2016vrs} do not 
observe any threshold. The reason behind this may be worth further investigation.   From a physical viewpoint, the
existence of the threshold itself receives further evidence, which is also due to the computation of new observables. A theoretical coherent scenario remains however missing. Finally, the results on the topological susceptibility open the way to a new computation of the limits and uncertainties  of the post-inflationary axion mass, and, for the first time, of the axion potential around $T_c$. We hope to return to these points in a future publication. 

\section{Acknowledgements}

The authors thank  Gernot Akemann, Claudio Bonati,  Kenji Fukushima, Robert Pisarski,  Giancarlo Rossi, Kalman Szabo and Giovanni Villadoro  for valuable discussions. 

The work has been carried out using the open-source packages tmLQCD~\cite{Jansen:2009xp,Abdel-Rehim:2013wba,Deuzeman:2013xaa,Kostrzewa:2022hsv}, LEMON~\cite{Deuzeman:2011wz}, DD-$\alpha$AMG~\cite{Frommer:2013fsa,Alexandrou:2016izb,Bacchio:2017pcp,Alexandrou:2018wiv}, QPhiX~\cite{10.1007/978-3-319-46079-6_30,Schrock:2015gik} and QUDA~\cite{Clark:2009wm,Babich:2011np,Clark:2016rdz}, and 
the authors thank  B.~Kostrzewa for discussions and  help with the ETMC software. 

A.Yu.K is supported by ERC-MUON-101054515.
M.P.L is partially supported  by the COST (European Cooperation in Science and Technology) Action COSMIC WISPers CA21106.
The authors are grateful to ECT* for support for the Workshop “New developments in studies of the QCD phase diagram” during which this work has been developed.
Numerical simulations have been carried out using computing resources of CINECA, grants SIM2024  and IsCa8.

\bibliographystyle{JHEP}
\bibliography{refs}

\end{document}